\documentclass[12pt,showpacs,amsmath,amssymb,aps,floats,floatfix,nofootinbib]{revtex4}%
\usepackage{epsfig}
\usepackage{dcolumn}
\usepackage{bm}
\usepackage{amsmath}
\usepackage{amsfonts}
\usepackage{amssymb}
\usepackage{graphicx,subfigure}
\usepackage{color}
\usepackage{latexsym}
\usepackage{slashed} 

\usepackage{pstricks}
\usepackage{indentfirst}
\usepackage{mathrsfs}
\usepackage{multirow}
\usepackage{epsfig,subfigure,psfrag}

\def\iab{\mathrm{ab}^{-1}} 
\def\TeV{\mathrm{TeV}}     
\def\GeV{\mathrm{GeV}}     
\def\pT{p_\mathrm{T}} 
\def\missET{\slashed E_\mathrm{T}} 

\makeatother

\allowdisplaybreaks 

\begin{document}

\title{Searches for dark matter signals in simplified models at future hadron colliders}

\author{Qian-Fei Xiang}
\author{Xiao-Jun Bi}
\author{Peng-Fei Yin}
\author{Zhao-Huan Yu}
\affiliation{Key Laboratory of Particle Astrophysics,
Institute of High Energy Physics, Chinese Academy of Sciences,
Beijing 100049, China}

\begin{abstract}
We study the prospect of dark matter (DM) searches in the monojet channel at future $pp$ colliders with center-of-mass energies of 33, 50, and 100~TeV. We consider a class of simplified models in which a vector boson connecting DM particles to quarks is introduced.
Comparing with studies in the effective field theory, the present
framework gives more reasonable
production rates and kinematics of the DM signatures.
We estimate the sensitivities of future colliders with an integrated
luminosity of $3~\iab$ to the DM-induced monojet signature and
show the parameter space that can be explored. The constraints from direct and indirect DM detection experiments are compared with the future collider sensitivities. We find that the future collider detection will be much more sensitive than the indirect detection for the vector interaction, and have better sensitivities than those of the direct detection by several orders of magnitude for the axial vector interaction.
\end{abstract}

\maketitle

\section{Introduction}

Observations of astrophysics and cosmology reveal that the main component of matter in the Universe is dark matter.
However, the nature of DM particles is still unclear. Although the standard model (SM) has achieved great success after the LHC discovery
of a $\sim 125$~GeV Higss boson, it cannot provide any suitable candidate for the cold DM. Therefore, the existence of DM gives a hint of new physics beyond the SM (BSM). The most popular and attractive candidates for DM are the so-called weakly interacting massive particles (WIMPs), which can be thermally produced in the early Universe and naturally give the correct observed DM relic density. Because of their weak interactions with SM particles, WIMPs with masses of $\sim \mathcal{O}(10^2)$~GeV are expected to be produced at high energy colliders. Therefore, searching for such DM particles is a very important task for future collider experiments.

DM particles can be produced by cascade decays of heavy new particles in new physics models at colliders. A typical example is the lightest neutralino in supersymmetric models, which predict a bunch of new particles that have not been discovered by LHC searches up to now.
Another possibility is that DM particles are directly pair-produced in collisions of SM particles. In this case, DM signatures have been widely studied in the context of effective field theory (EFT)~\cite{Beltran:2010ww,Fan:2010gt,Cao:2009uw,Goodman:2010yf,Goodman:2010ku,Cheung:2010ua,Zheng:2010js,Cheung:2011nt,Yu:2011by,Fitzpatrick:2012ix}, which provides a ``model-independent'' way to quantitatively compare the sensitivity among collider, direct, and indirect detection experiments. Since DM particles escape from the detector without energy deposit, an additional energetic jet or photon is required to reconstruct the signature with missing transverse energy ($\missET$). The latest ATLAS and CMS searches have not found such so-called ``monojet'' and ``monophoton'' signatures induced by DM particles, and hence set constraints on the energy scales of effective contact operators describing DM interactions with SM particles in the EFT~\cite{Aad:2015zva,Aad:2014tda,Khachatryan:2014rra,Khachatryan:2014rwa}.

As long as the mediators interacting with both DM and SM particles are so heavy that they can be integrated out, the EFT approach is valid. However, it has many shortages and would break down in many cases, as pointed out in several recent works~\cite{Bai:2010hh,Fox:2011pm,Shoemaker:2011vi,Buchmueller:2013dya,Busoni:2013lha,Busoni:2014sya,Endo:2014mja,Busoni:2014haa,Buchmueller:2014yoa}.
It is well known that the EFT fails when the typical momentum transfer involved in the reaction is comparable to the mediator mass. This means that in order to safely use the EFT, the scale of BSM physics should be much higher than the collision energy, otherwise mediators may be directly produced in collisions. Besides, the EFT approach is invalid for a mediator with a width comparable to or larger than its mass~\cite{Fox:2011pm,Buchmueller:2014yoa}. Furthermore, unitary and perturbativity conditions could also set constraints on the validity of the EFT~\cite{Shoemaker:2011vi}.

In a more appropriate approach, so-called ``simplified models'', mediators with moderate masses are introduced to connect DM particles to SM particles. In principle, the simplified model approach can be mapped into the EFT approach, as studied in the limit of heavy mediators~\cite{Goodman:2010yf,Goodman:2010ku,Zheng:2010js,Yu:2011by}. On the other hand, it can be realized as particular cases of UV complete models, and has been widely used in supersymmetry studies (see e.g. Refs.~\cite{Cohen:2013xda,Kraml:2013mwa,Papucci:2014rja}). For a mediator with a moderate mass, full kinematics and topologies of DM signatures at colliders can be studied in details. Furthermore, collider constraints and reaches could also be easily compared with those from direct and indirect detection experiments in specific simplified models~\cite{Buchmueller:2013dya,Lebedev:2014bba,Buchmueller:2014yoa,Abdallah:2014hon,Berlin:2014tja,Yu:2014mfa,Busoni:2014gta,Hooper:2014fda,Alves:2015pea,Chen:2015tia}.

In this work, we study a class of minimal simplified models involving a DM particle and a neutral spin-1 mediator $Z'$. Each model is characterized by the following parameters: the mediator mass $m_{Z'}$, the DM particle mass $m_\chi$, the $Z'$ coupling to the DM particle $g_\chi$, and the $Z'$ couplings to quarks $g_q$. Recent 7 and 8~TeV LHC results can be used to set limits on the parameter space of these simplified models~\cite{Buchmueller:2014yoa,Busoni:2014gta}. It is expected that LHC searches will set more stringent constraints with larger integrated luminosity at its designed energy $13-14~\TeV$ in the next few years.
In this work, we investigate the prospects of exploring DM signals in the
framework of simplified models with $Z'$ at future $pp$ colliders.
Three collision energies of 33, 50, and 100~TeV for the very large hadron collider
and the super proton-proton collider are considered (for recent phenomenology studies, see
Refs.~\cite{Anderson:2013ida,Fowlie:2014awa,Wen:2014mha,Anchordoqui:2014wha,Alva:2014gxa}).
We study the sensitivity of future colliders to the monojet signatures and set constraints on the parameter space. It is expected that these future colliders will be powerful to detect DM particles and mediators with masses of $\mathcal{O}(\TeV)$. Finally, we compare the reaches of future collider searches with results from direct and indirect detection experiments.

This paper is organized as follows. In Sec.~II, we describe the minimal simplified DM models containing a DM particle and a vector mediator.
In Sec.~III, we discuss the kinematics of the monojet signature and estimate the sensitivity to these models at future colliders.
In Sec.~IV, we present constraints and future reaches of direct and indirect searches, and then compare them with the capability of future collider.
Finally, we conclude in Sec.~V.

\section{Minimal Simplified Dark Matter Models}

In this section, we describe a class of minimal simplified models, in which a spin-1 mediator $Z'$ connecting the DM particle~($\chi$) to quarks is introduced.
The DM particle is assumed to be a singlet under SM gauge symmetries and only couples to $Z'$.
Here we are not going to enumerate all the possible sets of particle spins and Lorentz structures; we only focus on two typical cases where the DM particle is  either a Dirac fermion or a complex scalar.
For the fermionic DM case, we consider
DM particles and quarks have a vector or axial vector interaction with $Z'$, which reads
\begin{eqnarray}
  \mathcal{L}_\mathrm{FV} & = & \sum_q g_q Z'_\mu \bar{q} \gamma^\mu q
     + g_{\chi} Z'_\mu \bar{\chi} \gamma^\mu \chi, \\
  \mathcal{L}_\mathrm{FA} & = & \sum_q g_q Z'_\mu \bar{q} \gamma^\mu \gamma^5 q + g_{\chi} Z'_\mu \bar{\chi} \gamma^\mu \gamma^5 \chi,
\end{eqnarray}
where the sums run over all quark flavors. For the scalar DM case, axial vector current cannot be constructed for the DM particle, thus we only consider the vector current interaction given by
\begin{equation}
\mathcal{L}_\mathrm{SV}  = \sum_q g_q Z'_\mu \bar{q} \gamma^\mu q
     + i g_\chi Z'_\mu [\chi^* \partial^\mu \chi - (\partial^\mu \chi^*) \chi].
\end{equation}
The subscripts FV and FA represent the fermionic DM case with vector and axial vector interactions, respectively, while SV represents the scalar DM case with vector interaction. Here we have assumed that all the interactions are renormalizable and CP invariant. These simplified models are typical, and their discussions can be easily extended to other models with other interactions in principle.

Free parameters in these models are $m_\chi$, $m_{Z'}$, $g_\chi$, and $g_q$. In general, the $Z'$ couplings  to quarks $g_q$ are flavor dependent, and should be determined by the details of underlying UV complete models.
For simplicity, we just assume $g_q$ is universal for all the quarks; such an assumption can provide an illuminating example of DM searches.

Comparing with $g_\chi$, $g_q$ would suffer from additional limits from dijet resonance searches for $q\bar{q}\rightarrow Z' \rightarrow q\bar{q}$ at hadron colliders~\cite{An:2012va,Yu:2013wta}. Especially, it is expected that future colliders will have strong capability to search for dijet resonances due to their high energies and luminosities. Combining the results from dijet and monojet searches, constraints on $g_q$ and $g_\chi$ can be simultaneously set. In this work, we only focus on monojet searches and simply set $g_q=g_\chi$ in order to further reduce the number of free parameters. Since the DM production rate is almost independent from the $Z'$ width $\Gamma_{Z'}$ and just sensitive to the product of $g_q^2 $ and $g_\chi^2$ in most of the parameter space except for the resonance region, the discussions on monojet searches under the assumption $g_q=g_\chi$ can be easily extended to the $g_q\neq g_\chi$ case in these regions.

$\Gamma_{Z'}$ would play an important role in the resonance region. Since $\Gamma_{Z'}$ depends on the sum of the terms proportional to $g_q^2$ and $g_\chi^2$, $\Gamma_{Z'}$ and $g_q^2 g_\chi^2 $ can be simultaneously treated as free parameters instead of $g_q$ and $g_\chi$. The detailed discussions on the effect of $\Gamma_{Z'}$ and the $g_q\neq g_\chi$ case in DM searches can be found in several works~\cite{Fox:2011pm,Buchmueller:2013dya, An:2012va,Buckley:2014fba}.
The $Z'$ width can be expressed as
\begin{equation}
   \Gamma_{Z'} = \Gamma(Z'\to \chi \bar\chi/\chi\chi^*) \Theta(m_{Z'}-2 m_{\chi})
     +\sum_q c_q  \Gamma(Z'\to q\bar{q}) \Theta(m_{Z'}-2 m_{q}), \label{eq:width_v}
\end{equation}
where the step function means that the particular $Z'$ decay channel opens when it is allowed by kinematics.
The color factor $c_q = 3$.
Partial widths contributed by individual decay channels are
\begin{align}
\Gamma_\mathrm{FV} (Z' \to q\bar{q}) &=\frac{m_{Z'}}{12 \pi}
   g_q^2 \xi_q \left(1+\frac{2 m_q^2}{m_{Z'}^2}\right),
&\Gamma_\mathrm{FV} (Z' \to \chi\bar{\chi})  &=  \frac{m_{Z'}}{12 \pi} g_\chi^2 \xi_\chi \left(1+\frac{2 m_\chi^2}{m_{Z'}^2}\right);
  \\
\Gamma_\mathrm{FA}  (Z' \to q\bar{q}) &=\frac{m_{Z'}}{12 \pi}
  g_q^2 \xi_q^3,
&\Gamma_\mathrm{FA}  (Z' \to \chi\bar{\chi}) &=\frac{m_{Z'}}{12 \pi}
  g_\chi^2 \xi_\chi^3; \\
\Gamma_\mathrm{SV} (Z' \to q\bar{q}) &= \Gamma_\mathrm{FV}(Z' \to \bar{q}q),
&\Gamma_\mathrm{SV} (Z' \to \chi\chi^*) &=\frac{m_{Z'}}{12 \pi}
  \frac{g_\chi^2}{4}\xi_\chi^3. \label{eq:width_sv}
\end{align}
Here $\xi_{f} \equiv \sqrt{1-4 m_f^2/m_{Z'}^2}$ and can be regarded as the particle velocity in the $Z'$ rest frame.

If the $Z'$ width is larger than the $Z'$ mass, it would be questionable to treat $Z'$ as a particle.
Therefore, the couplings $g_\chi$ and $g_q$ should not be too large.
In view of the fact that $\xi_{q}<1$ and $\xi_q(1+2 m_q^2/m_{Z'}^2)<1$,
we have $\Gamma_{Z'} < (g_\chi^2+\sum_q c_q g_q^2)\,m_{Z'}/(12 \pi)$
for the fermionic DM case.
Consequently, the requirement $\Gamma_{Z'} < m_{Z'}$ gives
\begin{equation}
g_\chi^2 + \sum_q c_q g_q^2 < 12 \pi.
\end{equation}
Under the assumption of $g_\chi=g_q$, we can get a rough limit of $g_q < 1.4$. The limit on $g_q$ for the scalar DM case is similar. Moreover, the perturbativity condition also requires that the couplings should not be too large.
For these reasons, below we only consider the parameter space with $g_q(g_\chi)\le 1$.

\section{Monojet signature at future colliders}

If DM particles are produced in pair via high energy $pp$ collisions, an additional energetic object, such as a jet, photon, charged lepton, or $Z$ boson, is required to reconstruct $\missET$ and trigger the event.
In the following, we investigate the monojet signature $pp\to Z^{\prime (*)}(\to \chi\bar\chi/\chi\chi^*)+\mathrm{jets}$ at future $pp$ colliders with center-of-mass energies of 33, 50, and 100~TeV.
Background and signal samples at the parton level are generated by \texttt{MadGraph~5}~\cite{Alwall:2014hca}, to which the simplified models are
added through \texttt{FeynRules~2}~\cite{Alloul:2013bka}. We use \texttt{PYTHIA~6}~\cite{Sjostrand:2006za} to deal with particle decay, parton shower, and hadronization processes, and adopt
the MLM matching scheme. \texttt{Delphes~3}~\cite{deFavereau:2013fsa} is utilized to carry out a fast detector simulation. Jets are reconstructed using the anti-$k_T$ algorithm~\cite{Cacciari:2008gp} with a distance parameter $R=0.4$. For realistic search strategies using a future detector designed with higher efficiency and resolution, results in this work are expected to be improved.

The dominant SM backgrounds are $Z(\to \bar{\nu}\nu)+\mathrm{jets}$ and $W(\to l\nu)+\mathrm{jets}$. For the $W(\to l\nu)$+jets process, charged leptons may be clustered into a nearby jet or undetected when they close to the beam pipe. On the other hand, the $Z(\to \bar{\nu}\nu) +\mathrm{jets}$ background is irreducible. We optimize selection criteria to efficiently suppress backgrounds and maximize the statistical significance.

There should be at least an energetic jet in the final states. The leading jet $j_1$ is required to have $|\eta(j_1)|< 2.4$ and $\pT(j_1) > 1.6/1.8/2.6~\TeV$ for $\sqrt{s}=33/50/100~\TeV$.
Events with more than two jets with $\pT>100$GeV and $|\eta|< 4$ are rejected.
The DM production process may involve more than one jet from initial state radiation.
In order to select more signal events, a second jet ($j_2$) is allowed if it satisfies the condition $\Delta\phi(j_1,j_2) < 2.5$. The cut on $\Delta\phi(j_1,j_2)$ is necessary to suppress the QCD multijet background, where large fake $\missET$ may come from inefficient measurement of one of the jets. Furthermore, in order to reduce other backgrounds, such as $W(\to l\nu)+\mathrm{jets}$, $Z(\to \ell^+\ell^-)+\mathrm{jets}$, and $t\bar{t}+\mathrm{jets}$ with leptonic top decays, the
events containing isolated electrons, muons, taus, or photons with $\pT>20~\GeV$ and $|\eta|< 2.5$ are discarded.

\begin{figure}[!htbp]
\centering
\subfigure[~$\sqrt{s}=33$~TeV]{
\includegraphics[width=0.45\textwidth]{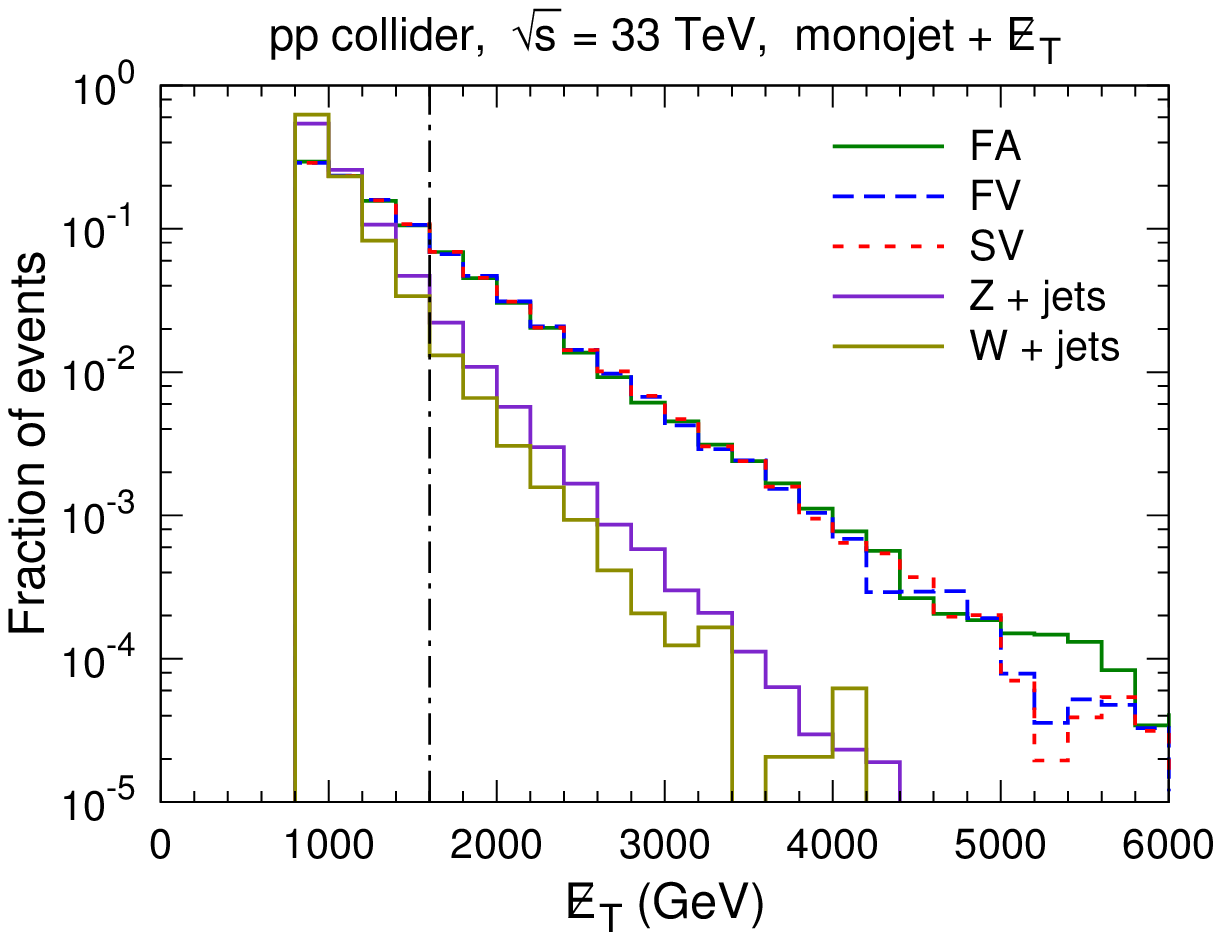}}
\subfigure[~$\sqrt{s}=50$~TeV]{
\includegraphics[width=0.45\textwidth]{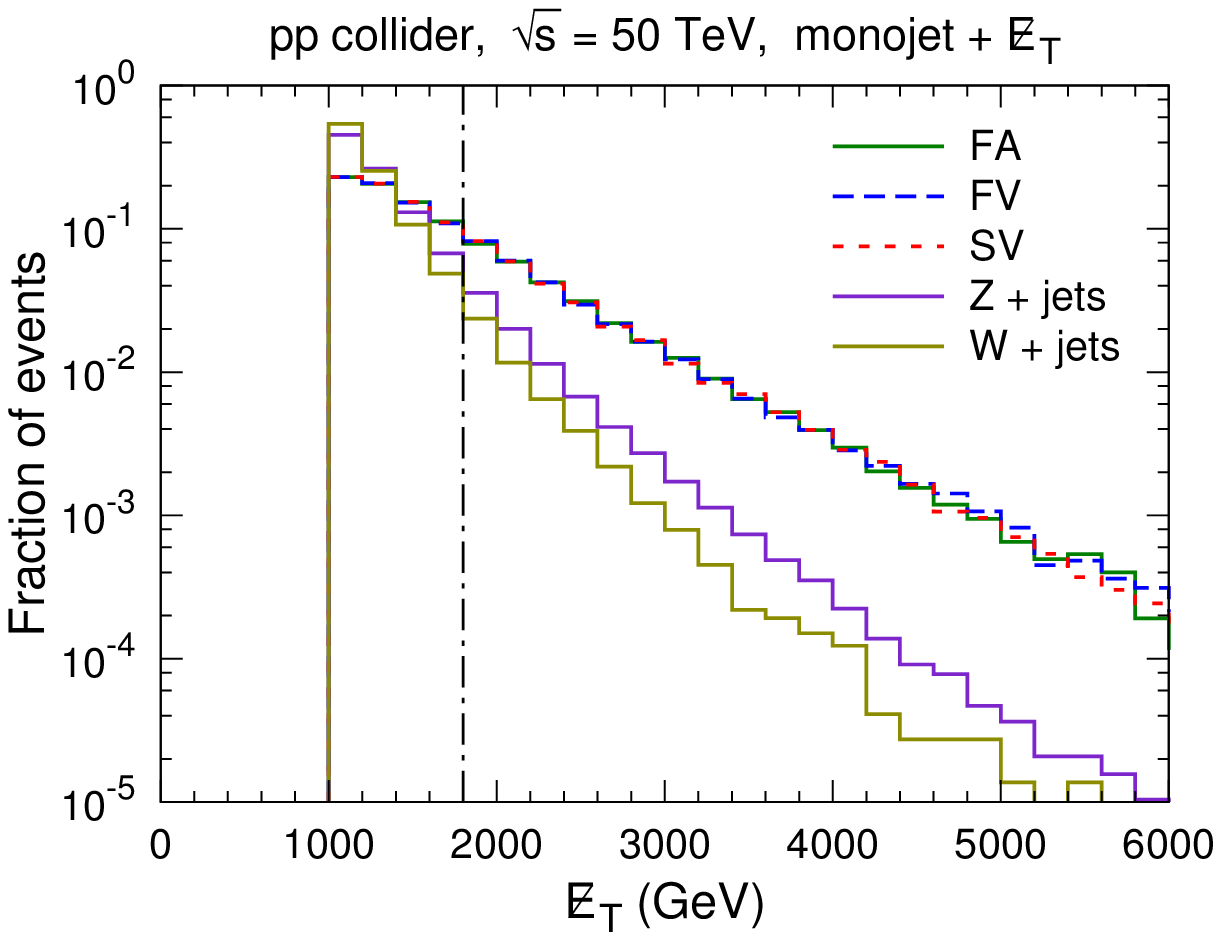}}
\subfigure[~$\sqrt{s}=100$~TeV]{
\includegraphics[width=0.45\textwidth]{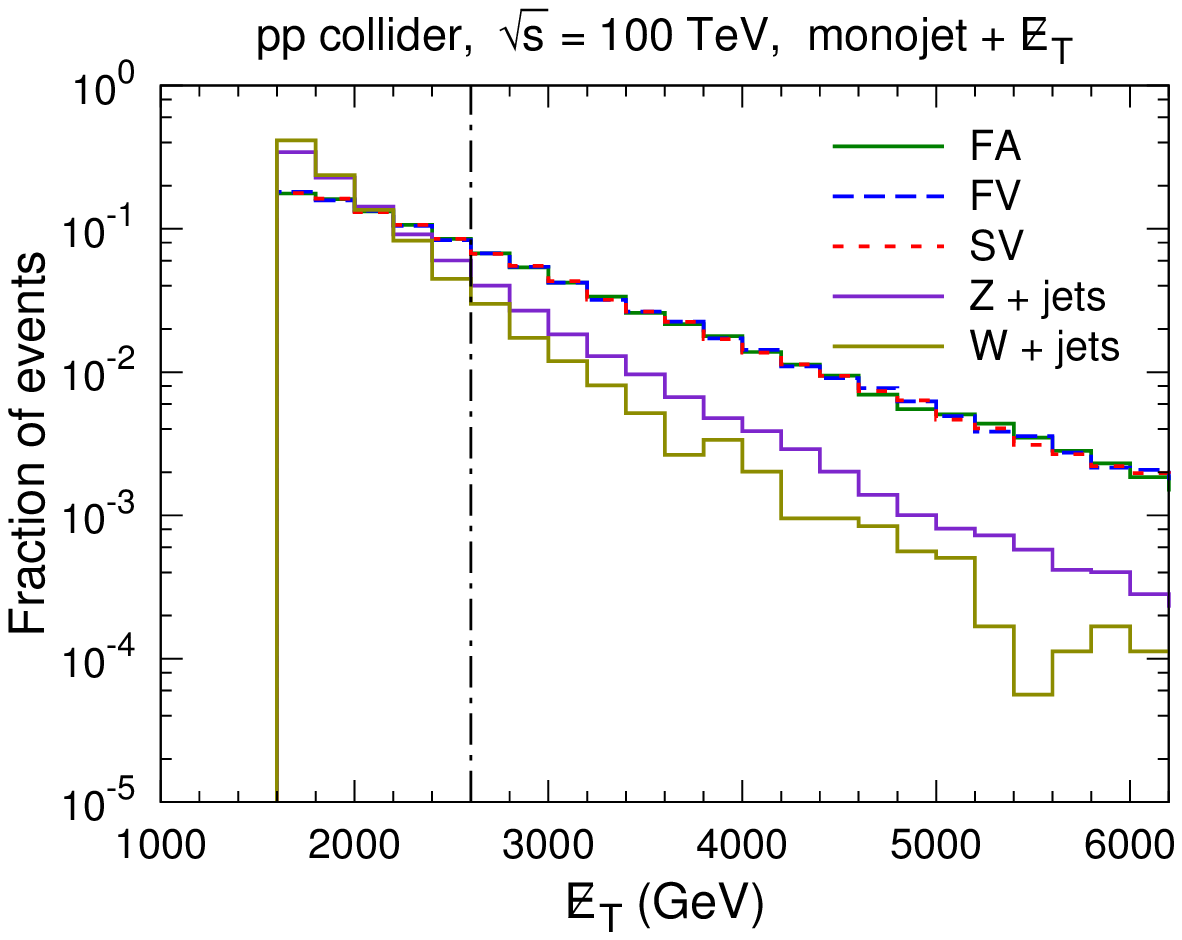}}
\caption{Normalized $\missET$ distributions for signal and background samples at $\sqrt{s}$ = 33~TeV~(a), 50~TeV~(b), and 100~TeV~(c) with thresholds of 800, 1000, and 1600~GeV, respectively. For DM production signals $pp\to Z^{\prime (*)}(\to \chi\bar\chi/\chi\chi^*)+\mathrm{jets}$, $g_q=g_\chi=0.1$, $m_\chi = 1~\TeV$, and $m_{Z'} = 5~\TeV$ are assumed.
The dot-dashed vertical lines denote the locations of the $\missET$ cuts used in the analysis.}
\label{fig:Missing_energy}
\end{figure}

The cut condition on $\missET$ is crucial for DM searches. In Fig.~\ref{fig:Missing_energy}, we show the normalized $\missET$ distributions of signals and backgrounds, assuming $g_q=g_\chi=0.1$, $m_\chi = 1~\mathrm{TeV}$ and $m_{Z'} = 5~\mathrm{TeV}$.
The distributions become harder as the collision energy increases.
The slopes of the distributions are mainly determined by the energy scales involved in collisions.
Since here $m_{Z'}$ and $m_\chi$ are chosen to be much larger than $m_Z$ and $m_W$, signal distributions are always harder than background distributions.
The distributions for three simplified models are similar, implying that spins and Lorentz structures play minor roles in high energy collisions.
In order to optimize the statistical significance, we choose the cut condition as $\missET > 1.6/1.8/2.6~\TeV$ for $\sqrt{s}=33/50/100~\TeV$.

\begin{table}[!htbp]
\begin{center}
\setlength\tabcolsep{0.5em}
\caption{Cross sections (in fb) after cuts for backgrounds and signals. For DM production signals $pp\to Z^{\prime (*)}(\to \chi\bar\chi/\chi\chi^*)+\mathrm{jets}$, $g_q=g_\chi=0.1$, $m_\chi = 1~\TeV$, and $m_{Z'} = 5~\TeV$ are assumed for all the three simplified models.
\label{tab:cs}}
\vspace*{1em}
\begin{tabular}{ccccccc}
\hline\hline
$\sqrt{s}$   & $W(\to l\nu)+\mathrm{jets}$ & $Z(\to \bar{\nu}\nu)+\mathrm{jets}$ & FV & FA & SV \\
\hline
33 TeV & $8.179 \times 10^1   $  & $1.948 \times 10^2$
                        & $3.043\times 10^{-2}$ & $2.399 \times 10^{-2}$  & $6.133 \times 10^{-3}$ \\
50 TeV & $6.991 \times 10^1   $  & $1.816 \times 10^2$
                        & $9.037\times 10^{-2}$ & $7.054 \times 10^{-2}$  & $1.824 \times 10^{-2}$ \\
100 TeV & $3.475 \times 10^1   $  & $1.062 \times 10^2$
                        & $2.340\times 10^{-1}$ & $1.851 \times 10^{-1}$  & $4.735 \times 10^{-2}$ \\
\hline\hline
\end{tabular}
\end{center}
\end{table}

In Table~\ref{tab:cs}, we list the cross sections of backgrounds and signals after cuts at $\sqrt{s}=33$, 50, and 100~TeV, assuming $g_q=g_\chi=0.1$, $m_\chi = 1~\TeV$, and $m_{Z'} = 5~\TeV$.
Since we choose stricter cuts at higher collision energy, the cross sections of backgrounds decrease as $\sqrt{s}$ increases.
DM production rates in the FV and FA models are similar, implying that the influence of Lorentz structures is slight at high energy. Cross sections of scalar DM particles
are smaller than those of fermionic DM particles by a factor smaller than 10.
This is because scalar particles have less helicity states and their production cross section suffers a kinematic suppression in the angular distribution.

\begin{figure}[!htbp]
\centering
\subfigure[~Fermionic DM with vector $Z'$]{
\includegraphics[width=0.45\textwidth]{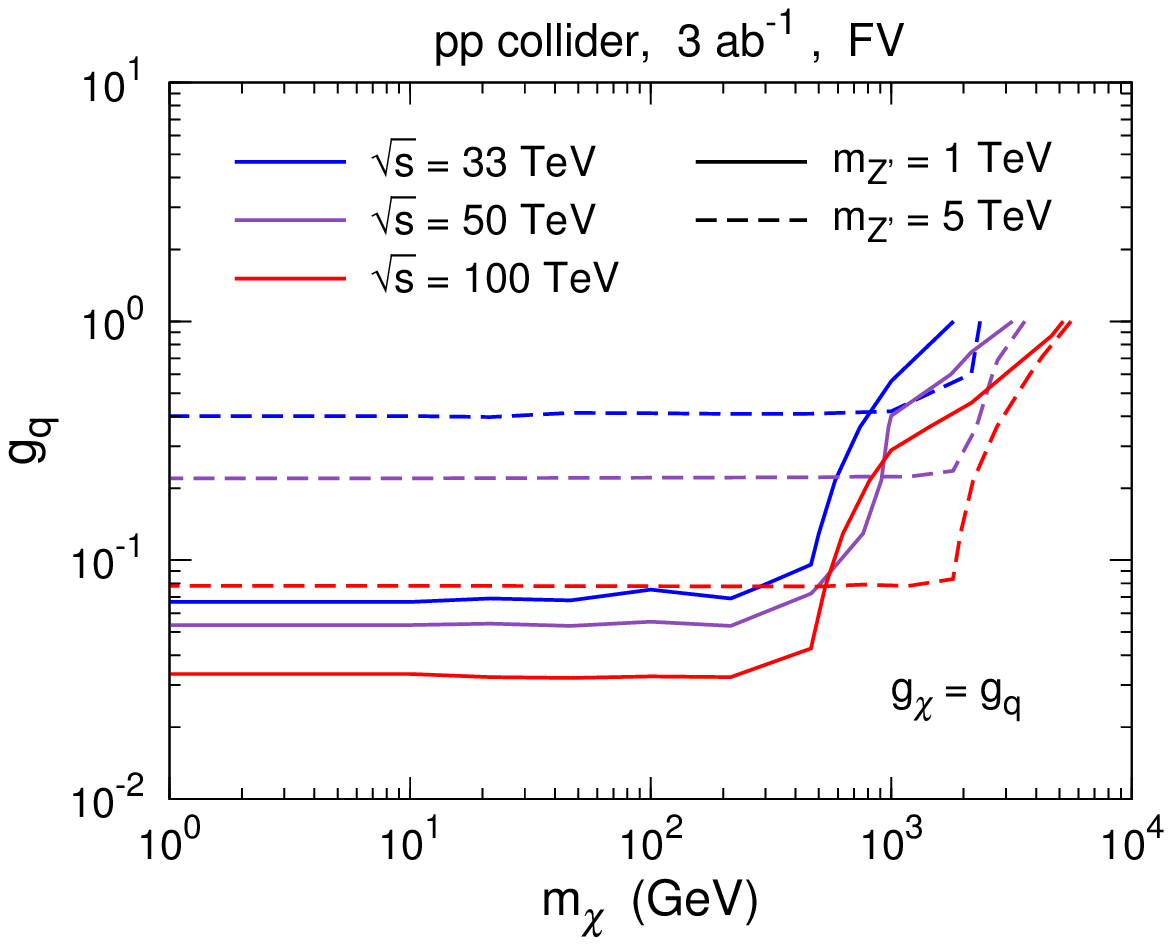}}
\subfigure[~Fermionic DM with axial vector $Z'$]{
\includegraphics[width=0.45\textwidth]{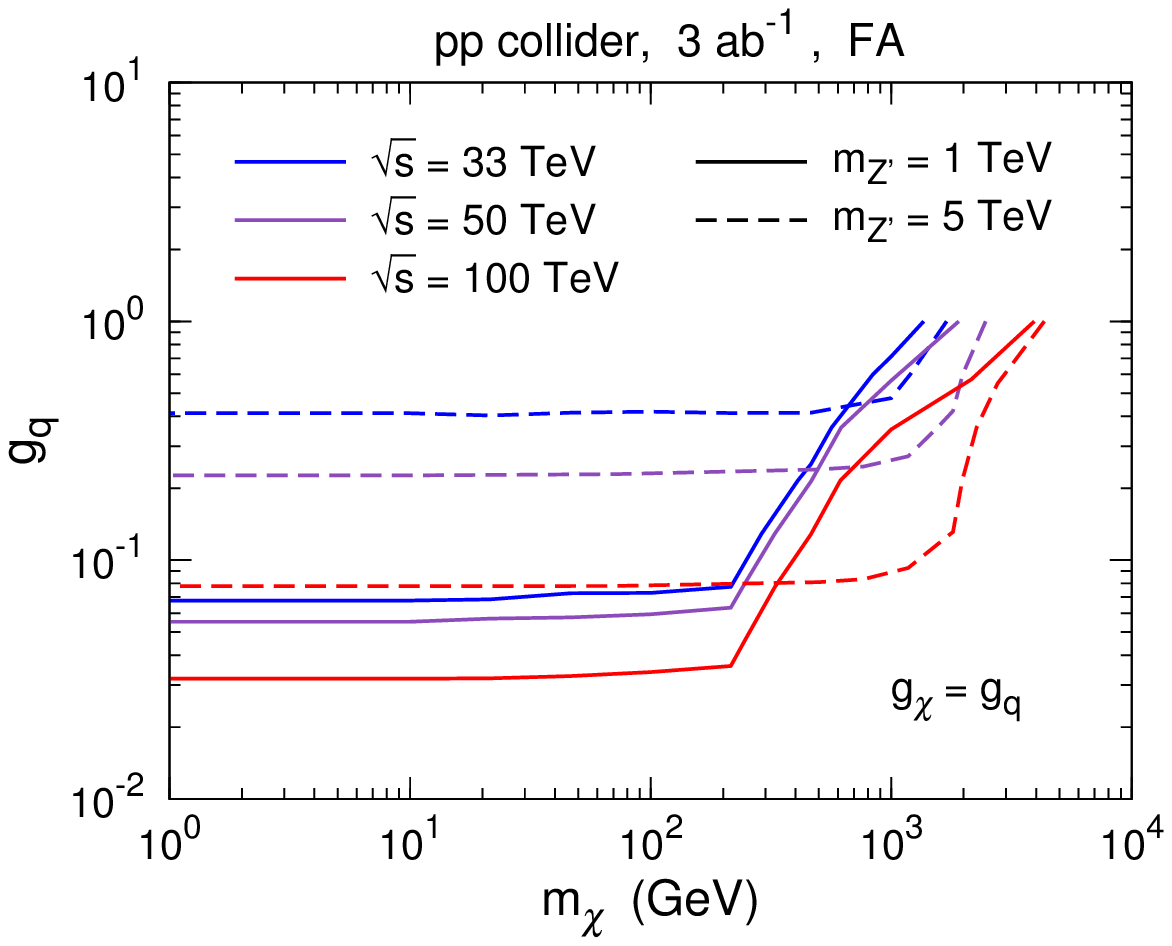}}
\subfigure[~Scalar DM with vector $Z'$]{
\includegraphics[width=0.45\textwidth]{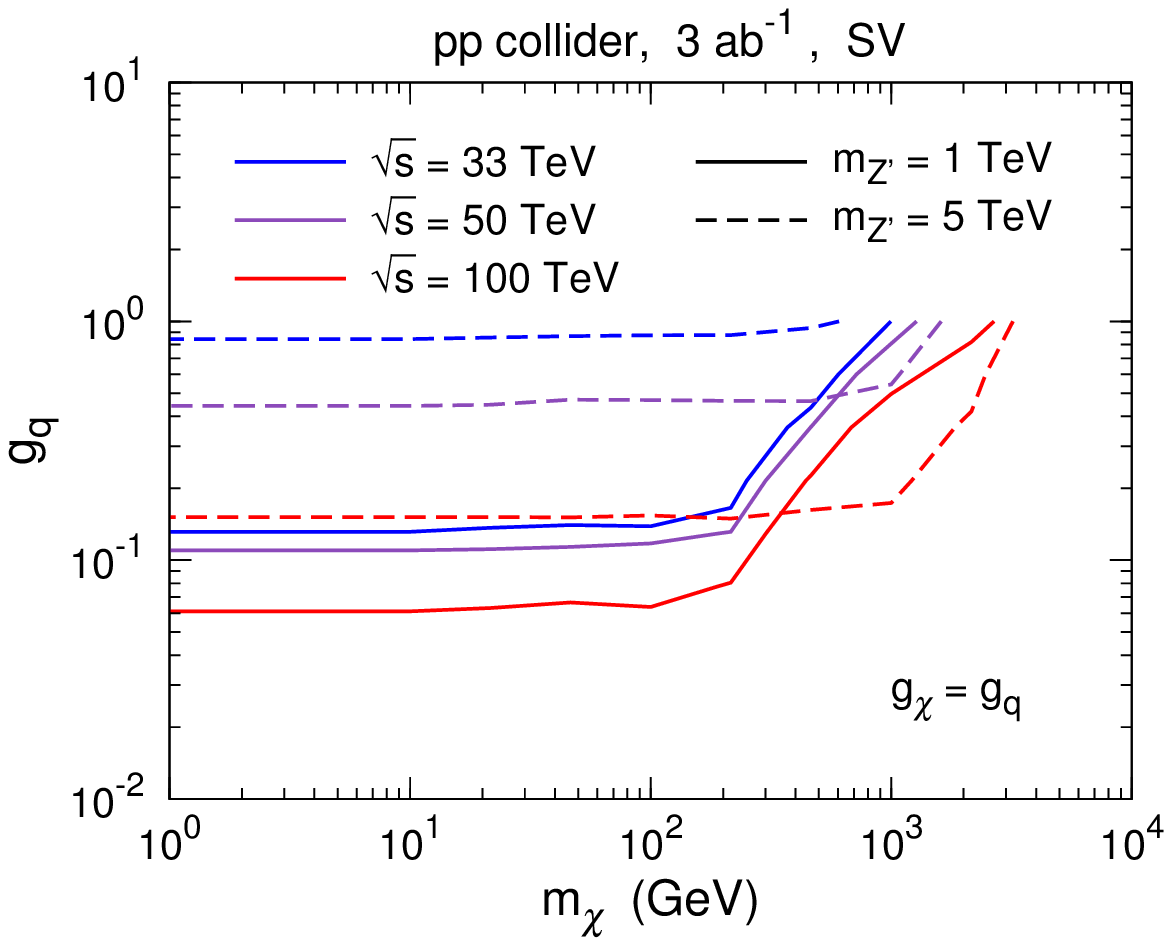}}
\caption{Estimated 90\% C.L. limits on the FV~(a), FA~(b), and SV~(c) models in the $m_{\chi}$-$g_q$ plane for the
$\mathrm{monojet}+\missET$ channel at $pp$ collides with $\sqrt{s}=33$~TeV (blue lines), 50~TeV (purple lines), and 100~TeV (red lines), assuming an integrated luminosity of $3~\iab$.
Solid and dashed lines correspond to $m_{Z'}=1$~TeV and 5~TeV, respectively. The regions
above the curves are expected to be excluded. We have assumed $g_\chi=g_q$.
}
\label{fig:gq_limits}
\end{figure}

Now we investigate the sensitivity to the simplified DM models at future $pp$ colliders.
The estimated $90\%$ C.L. exclusion limits in the $m_{\chi}$-$g_q$ plane for $m_{Z'}=1$ and 5~TeV are shown in Fig.~\ref{fig:gq_limits}.
The DM production cross section dramatically depends on whether the mediator $Z'$ is on-shell or not.
When $m_\chi < m_{Z'}/2$, $Z'$ can be on-shell produced and then decays into a pair
of DM particles. In this case the DM production cross section is resonantly enhanced and
proportional to $g_q^2 \mathrm{Br}(Z'\to\chi\chi)$ under the narrow
width approximation. For $m_{Z'} \gg 2 m_\chi$, the branching
ratio to DM particles $\mathrm{Br}(Z'\to\chi\chi)$ is almost a constant, and can be estimated as 1/19, 1/19, and 1/73 for the FV, FA, and SV models, respectively.
In this limit, the DM production cross section is irrelevant to $m_\chi$.
This explains the behaviors of the limits for $m_\chi < m_{Z'}/2$ in Fig.~\ref{fig:gq_limits}.
Note that $\mathrm{Br}(Z'\to\chi\chi)$ in the FV model almost equals to that in the FA model. This is why the limits for these two models are nearly the same. For the SV model,
$\mathrm{Br}(Z'\to\chi\chi)$ is smaller than those of fermionic DM models by a factor of $\sim 4$, hence the limits on $g_q$ is weaker by a factor of $\sim 2$.

The sensitivity drops quickly as $m_\chi$ becomes larger than $m_{Z'}/2$ and $Z'$ is off-shell produced. In this case, the DM production cross section is proportional to
$[g_q g_\chi/(Q^2 - m_{Z'}^2)]^2$, where $Q^2$ is the typical momentum
transfer to the DM particle pair.
As a result, for a large enough $m_\chi$, the collider search may be
more sensitive to a heavier on-shell $Z'$ than a lighter off-shell $Z'$, as shown in Fig.~\ref{fig:gq_limits}.

When $Z'$ is off-shell produced and $m_{Z'}^2 \ll Q^2$, the DM production cross section scales as
$(g_q g_\chi/Q^2)^2$, which is irrelevant to $m_{Z'}$. It can be found in
Fig.~\ref{fig:gq_limits} that when $m_\chi$ increases, the limits for
different $m_{Z'}$ become close to each other.
On the other hand, if $m_{Z'}^2 \gg Q^2$, the cross section is proportional to $(g_q g_\chi/m_{Z'}^2)^2$ and can be matched
to that in the EFT approach with an effective energy scale of
$\Lambda_\mathrm{eff}=m_{Z'}/\sqrt{g_q g_\chi}$.

\section{Comparison among different DM detection experiments}

DM signal features in different simplified models are quite similar at colliders, but may be very different in other DM detection experiments.
In this section, we will discuss the prospects of different kinds of DM searches and give a comparison among them. Firstly, we consider DM direct detection experiments, which search for nuclear recoil signatures induced by DM-nucleus scatterings. Since the typical DM velocity near the Earth is $\sim 10^{-3}$, the momentum transfer in the DM-nucleus scattering is $\sim \mathcal{O}(\mathrm{KeV})$, which is much lower than that at colliders. Thus the nonrelativistic limit is valid and different DM simplified models could induce distinct phenomenologies in the direct detection. For instance, vector and axial vector DM-quark interactions induce spin-independent (SI) and spin-dependent (SD) DM-nucleus scatterings, respectively. It is well known that the SI scattering cross section is coherently enhanced by the square of the nucleon number in the nucleus. On the other hand, the SD scattering signature has no coherent enhancement and depends on the particular spin property of the target nucleus.

\begin{figure}[!htbp]
\centering
\subfigure[~FV model, spin independent]{
\includegraphics[width=0.42\textwidth]{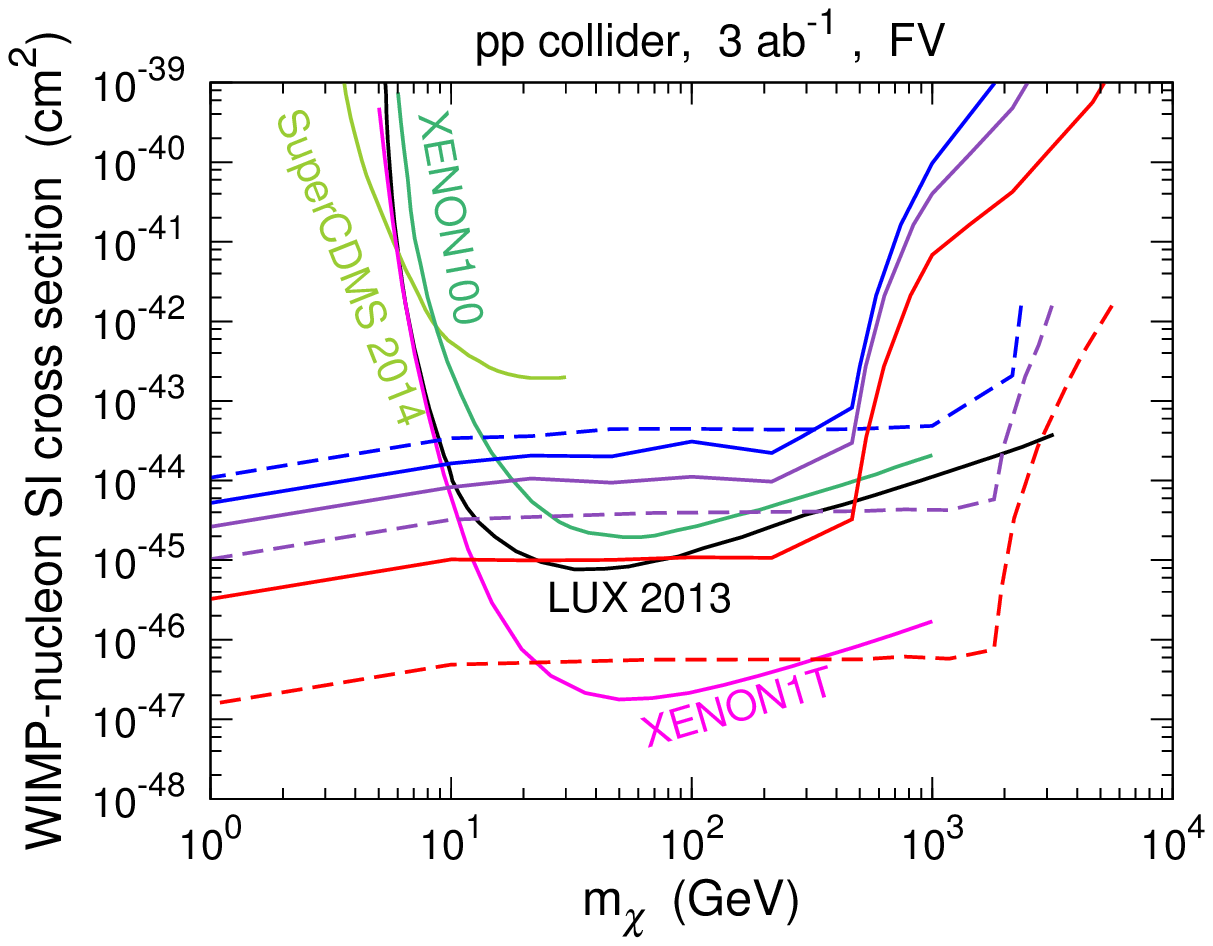}}
\subfigure[~FA model, spin dependent]{
\includegraphics[width=0.42\textwidth]{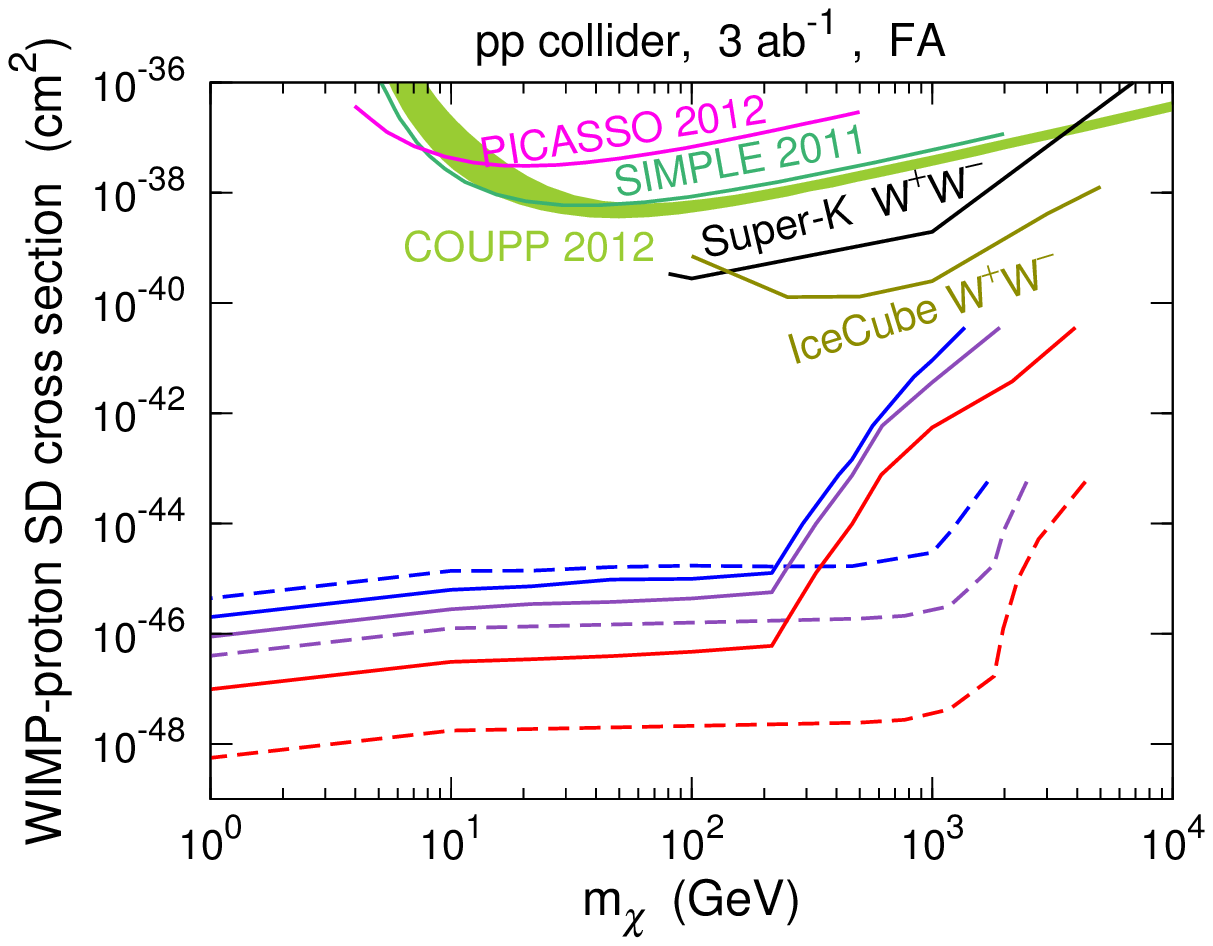}}
\subfigure[~SV model, spin independent]{
\includegraphics[width=0.42\textwidth]{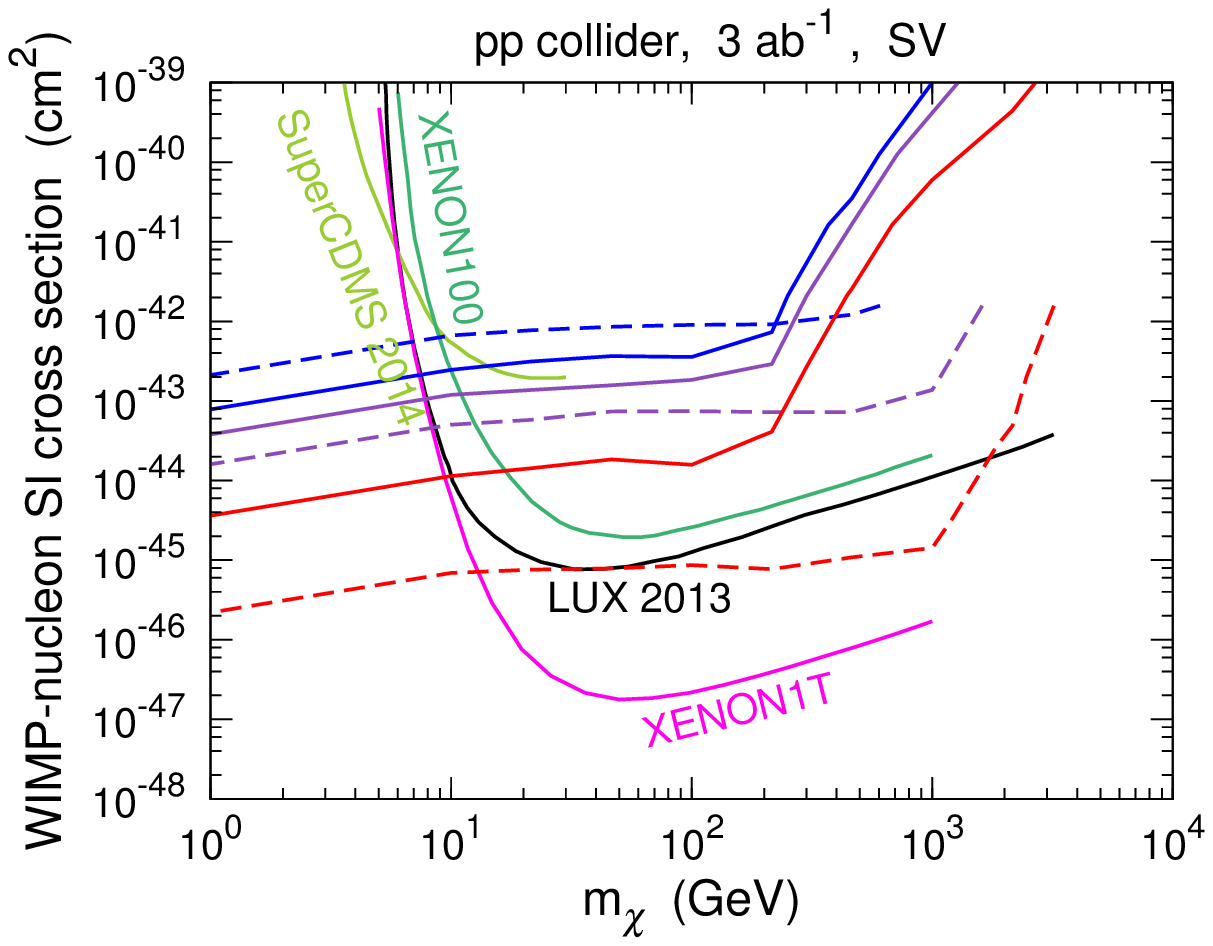}}
\caption{Estimated $90\%$ C.L. limits on the FV~(a), FA~(b), and SV~(c) models from the monojet searches in the $m_\chi$-$\sigma_{\chi N}$ plane.
Blue/purple/red lines correspond to $\sqrt{s}=33/50/100~\TeV$,
while solid (dashed) lines correspond to $m_{Z'}=1~(5)~\TeV$.
For the SI DM-nucleon scattering,
recent bounds from direct detection experiments XENON100~\cite{Aprile:2012nq},
LUX~\cite{Akerib:2013tjd}, and SuperCDMS~\cite{Agnese:2014aze}, and the expected reach of XENON1T~\cite{Aprile:2012zx} are shown. For the SD DM-nucleon scattering, current constraints from SIMPLE~\cite{Felizardo:2011uw}, PICASSO~\cite{Archambault:2012pm}, and COUPP~\cite{Behnke:2012ys} are shown, as well as the limits from neutrino detection experiments Super-K~\cite{Tanaka:2011uf} and IceCube~\cite{IceCube:2011aj}.}
\label{fig:DD}
\end{figure}

In order to compare reaches of the direct detection and the future collider detection, we show in Fig.~\ref{fig:DD} the translated 90\% C.L. limits from the monojet searches in the $m_\chi$-$\sigma_{\chi N}$ plane, where $\sigma_{\chi N}$ is the DM-nucleon scattering cross section. For the vector interaction, current direct detection experiments have set stringent
constraints on the SI signature. The collider detection would have a stronger capability to search for light DM in the region of $m_{\chi}\lesssim 10~\GeV$, where the direct detection dramatically lose the sensitivity due to the experimental threshold.
For the axial vector interaction, constraints from direct searches are
very weak. The monojet search will significantly improve current
direct detection limits by several orders of magnitude.

\begin{figure}[!htbp]
\centering
\includegraphics[width=0.48\textwidth]{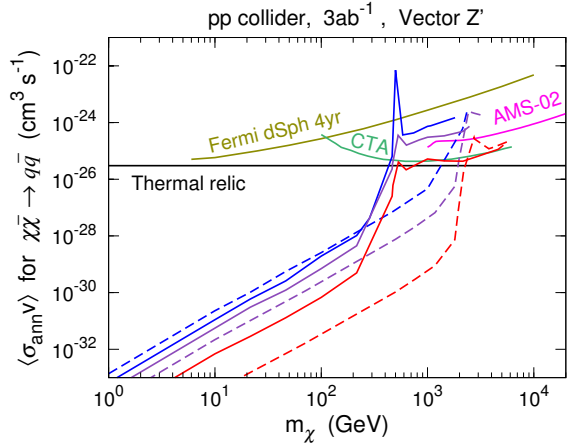}
\caption{Estimated 90\% C.L. limits on the FV model from the monojet searches in the $m_\chi$-$\langle \sigma_\mathrm{ann}v \rangle$ plane.
Blue/purple/red lines correspond to $\sqrt{s}=33/50/100~\TeV$,
while solid (dashed) lines correspond to $m_{Z'}=1~(5)~\TeV$.
For a comparison, also shown are the recent limit from the Fermi-LAT 4-year gamma-ray observations on the dwarf galaxies~\cite{Ackermann:2013yva} and the expected limits from the AMS-02 anti-proton detection for 20 years
and the CTA observation on the dwarf galaxy Segue~1 for 100 hours~\cite{Doro:2012xx}.
The horizontal line denotes the canonical thermal relic value $3\times 10^{-26}~\mathrm{cm}^3/\mathrm{s}$.}
\label{fig:sv}
\end{figure}

Indirect detection experiments search for high energy cosmic rays, gamma rays, and neutrinos induced by DM annihilations in the Galaxy and extragalactic objects. Signatures in these experiments depend on the thermally averaged DM annihilation cross section $\left< \sigma_\mathrm{ann}v \right>$
and DM density distributions in annihilating regions. In Fig.~\ref{fig:sv}, we demonstrate the translated 90\% C.L. limits from the
monojet searches in the $m_\chi$-$\langle \sigma_\mathrm{ann}v \rangle$
plane. The FV model leads to $s$-wave DM annihilations into quarks and can be explored by gamma-ray and cosmic-ray observations.
For $m_{\chi} \lesssim m_{Z'}/2$, future colliders could have a sensitivity better than indirect searches by several orders of magnitude.
In the FA model, DM annihilations into quarks are helicity suppressed in $s$-wave and hence depend on the DM velocity dispersion $\left< v^2 \right>$, which is typically $\sim\mathcal{O}(10^{-6})$ in the Galaxy.
In the SV model, DM annihilations into quarks are of $p$-wave in the leading order and also highly suppressed by $\left< v^2 \right>$.
Therefore, these two models cannot be explored via $q\bar{q}$ channels in indirect detection experiments.
Nonetheless, when $m_\chi > m_{Z'}$, DM particles can annihilate into $Z'$ pairs,
which would decay into quarks and give rise to detectable gamma-ray and cosmic-ray signals.

\begin{figure}[!htbp]
\centering
\subfigure[~FV model]{
\includegraphics[width=0.45\textwidth]{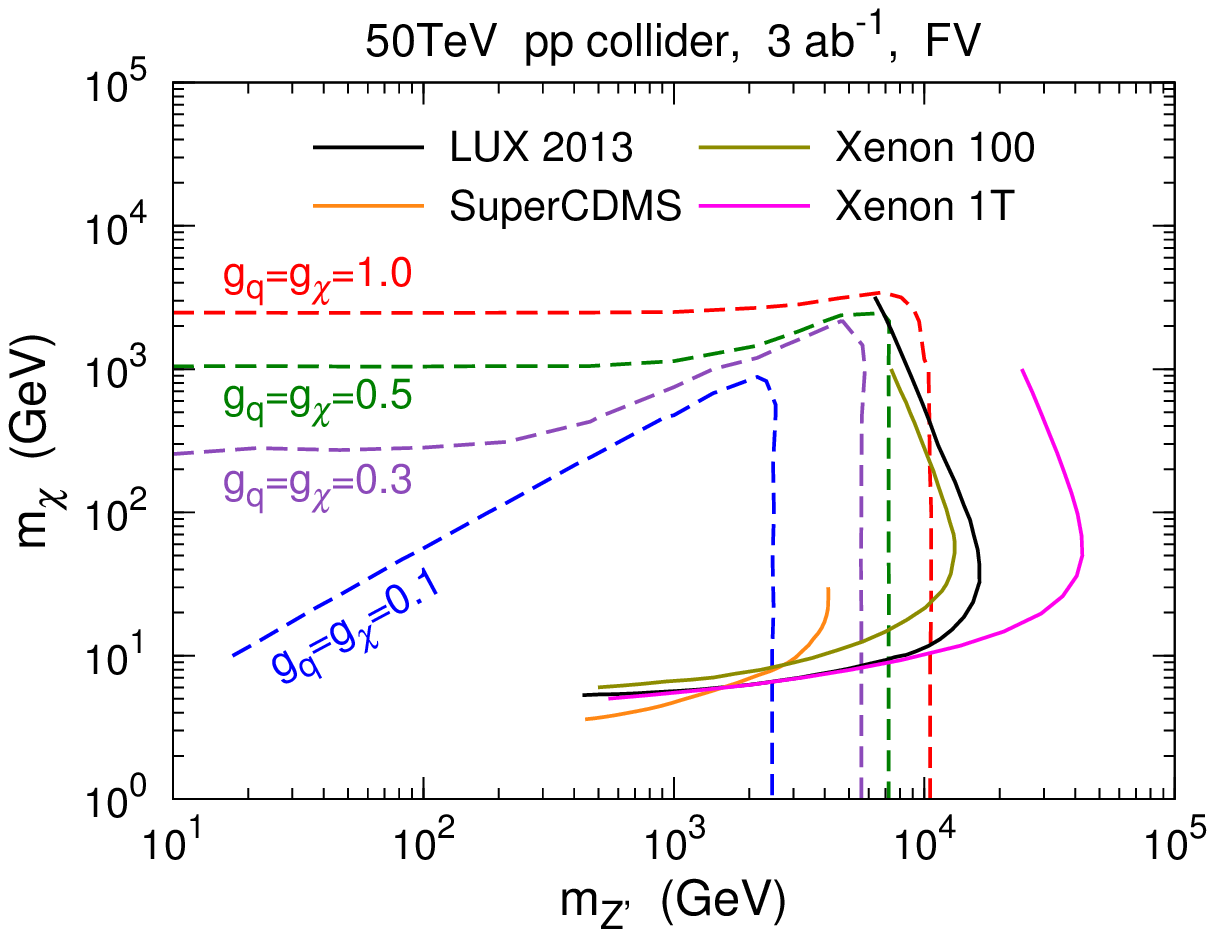}}
\subfigure[~FA model]{
\includegraphics[width=0.45\textwidth]{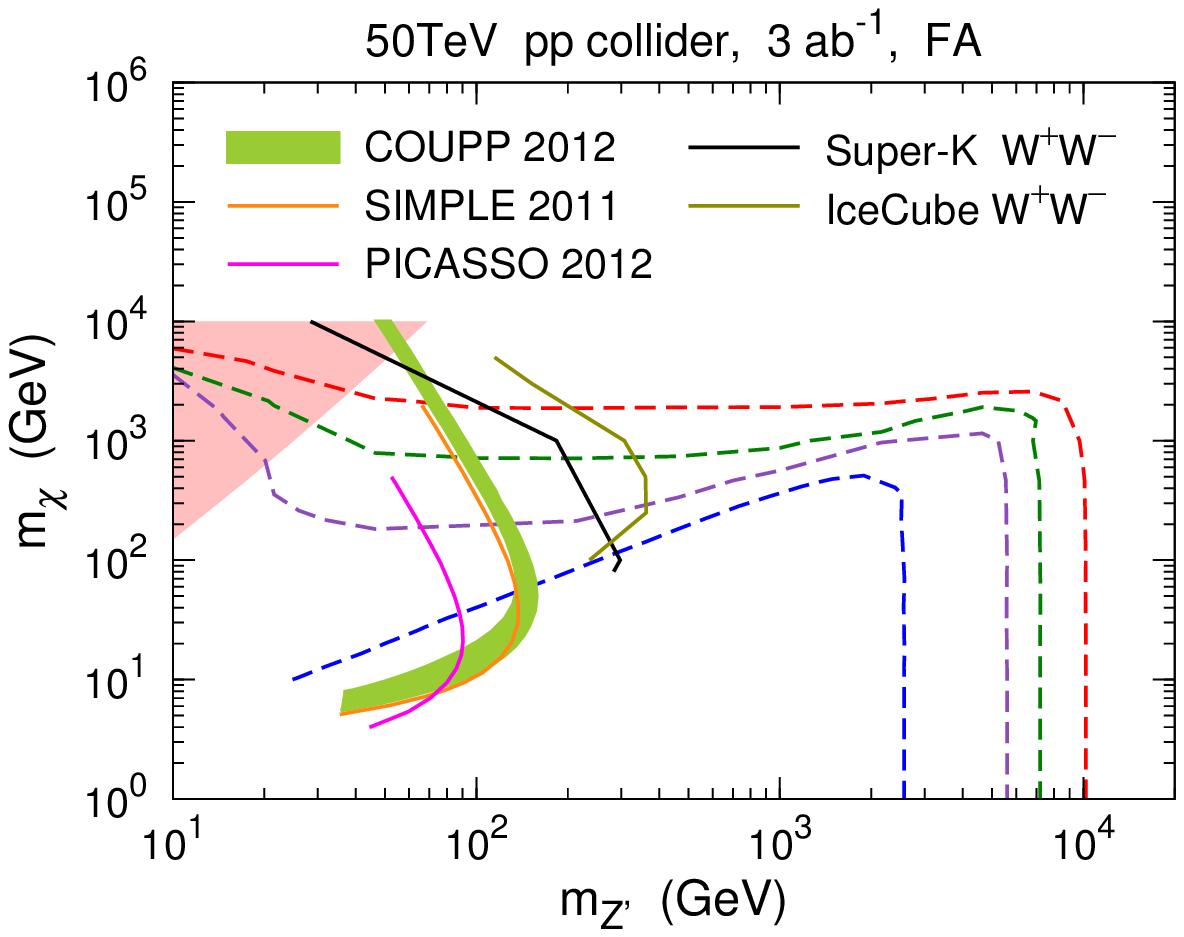}
\label{fig:unitary}}
\subfigure[~SV model]{
\includegraphics[width=0.45\textwidth]{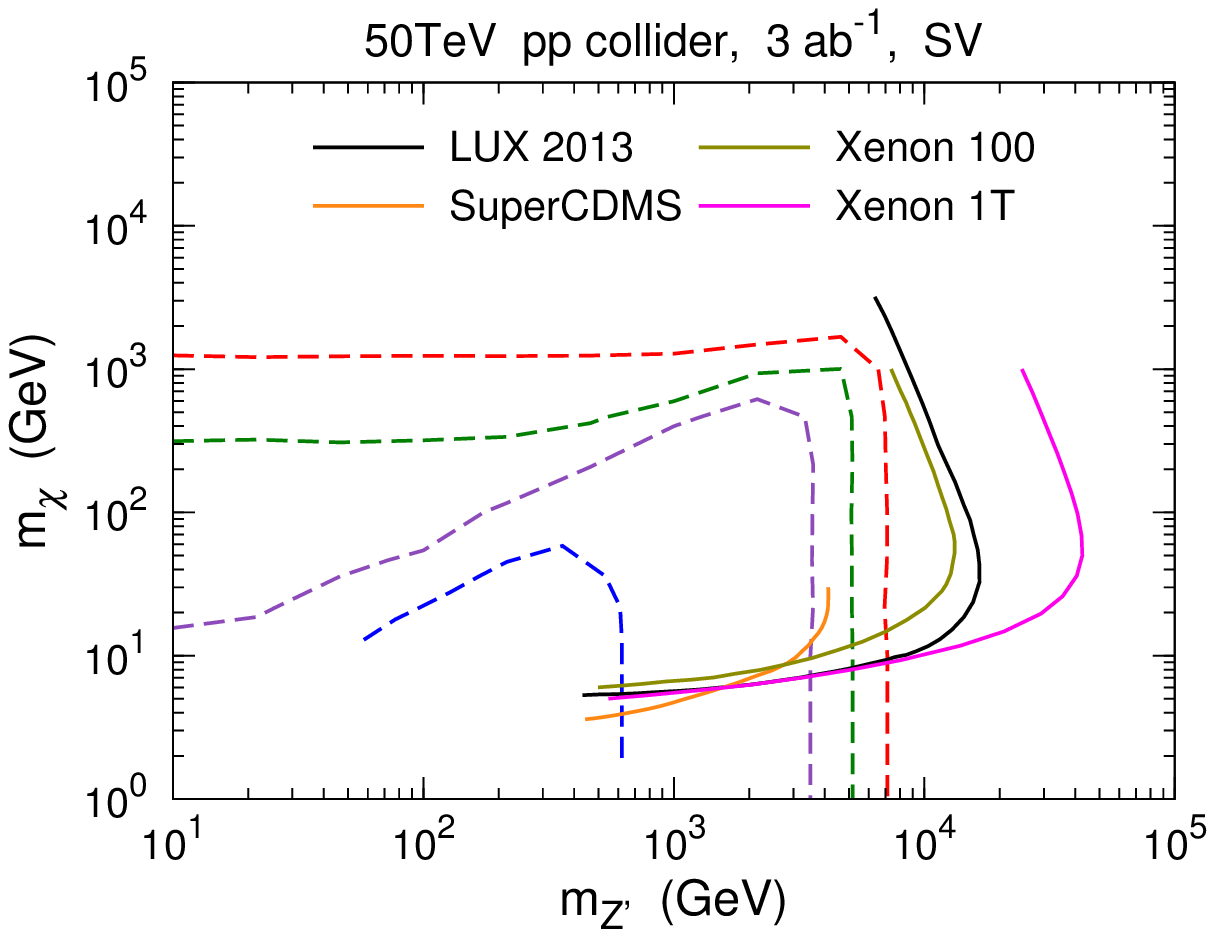}}
\caption{Estimated 90\% C.L. limits on the FV~(a), FA~(b), and SV~(c) models in the $m_{Z'}$-$m_\chi$ plane for the
$\mathrm{monojet}+\missET$ channel at a $pp$ collider with $\sqrt{s} = 50~\TeV$, in comparison with limits from direct searches. An integrated luminosity of $3~\iab$ is assumed.
Dashed blue/purple/green/red curves correspond to $g_q = g_\chi =0.1/0.3/0.5/1.0$.
The regions below these curves are expected to be excluded.
Limits from direct searches (solid lines) are converted assuming $g_q = g_\chi = 0.5$.
The unitarity violation region for $g_q=g_\chi=1$ in the FA model is indicated by light red color.}
\label{fig:50TeV_collider}
\end{figure}

In the above study, we have fixed $m_{Z'}$. Now we discuss the influence of the $Z'$ mass for DM searches.
Fig.~\ref{fig:50TeV_collider} shows the estimated 90\% C.L. limits in the
$m_{Z'}$-$m_\chi$ plane from the monojet search at a future $pp$ collider with fixed values of $g_q$. The collision energy and the integrated luminosity
are taken to be $50$~TeV and $3~\iab$, respectively.
The parameter space can be reasonably divided into two parts, $m_{\chi} < m_{Z'}/2$ and $m_{\chi} > m_{Z'}/2$.
As mentioned above, the DM production cross section is controlled by the factor $[g_q g_\chi/(Q^2 - m_{Z'}^2)]^2$ in the region of $m_{\chi} > m_{Z'}/2$,.
For $m_\chi \gg m_{Z'}$ and thus $Q^2 \gg m_{Z'}^2$, the DM production rate would be irrelevant to $ m_{Z'}$.
This explains the nearly horizontal segments of the estimated limits, except for the limits in the FA model when $m_{Z'}\lesssim 50~\GeV$, which will be discussed later.
For a weak coupling like $g_q=g_\chi=0.1$, the DM production rate with off-shell $Z'$ may be too low to detect, hence the limits are restricted to the region near $m_\chi=m_{Z'}/2$.
In the region of $m_{\chi} < m_{Z'}/2$, $Z'$ is on-shell produced and the limits are nearly vertical.
The maximum reach of $m_{Z'}$ only depends on the collision energy and is almost irrelevant to $m_\chi$.

In Fig.~\ref{fig:50TeV_collider}, we also map direct detection limits into the $m_{Z'}$-$m_\chi$ plane, assuming $g_q = g_\chi =0.5$.  For the FV and SV models, direct detection is much powerful than collider searches except for $m_\chi \lesssim 10~\GeV$. Hence the direct detection and collider detection can be complementary to each other in the parameter space. For the FA model, the collider detection will significantly improve the limits from the direct detection.

Since the FA model involves a massive vector boson coupling to non-conserved axial vector currents, it has a dangerous UV behavior.
When $m_{Z'}$ tends to zero, the DM production cross section is essentially proportional to $g_q^2 g_\chi^2 m_q^2 m_\chi^2/(m_{Z'}^4 Q^2)$ and would blow up.
Consequently, as shown in Fig.~\ref{fig:50TeV_collider}(b), the sensitivity for $g_q=g_\chi\ge 0.3$ is unusually improved for $m_{Z'}\lesssim 50~\mathrm{GeV}$.
In fact, the simplified model falls down in this case.
The problem could be solved in a UV complete model,
where more degrees of freedom are introduced and $m_{Z'}$ is no longer an arbitrary parameter but given by some physical mechanisms, such as spontaneous symmetry breaking. As $m_{Z'}$ tends to zero, the DM production process in the FA model may violate the unitarity condition.
In order to estimate the invalid region of this model, we derive a unitarity bound from the parton-level process $q\bar q\to\chi\bar\chi$ with an appropriate center-of-mass energy of the quark pair. The detailed derivation is described in Appendix~\ref{app:uni}.
For $g_q=g_\chi=1$, the unitary bound for $b\bar b\to\chi\bar\chi$ are shown in Fig.~\ref{fig:50TeV_collider}(b) by a region with light red color.
In this region, the FA model cannot be a correct description and the collider limits derived in this approach are meaningless.

\begin{figure}[!htbp]
\centering
\subfigure[~FV model]{
\includegraphics[width=0.45\textwidth]{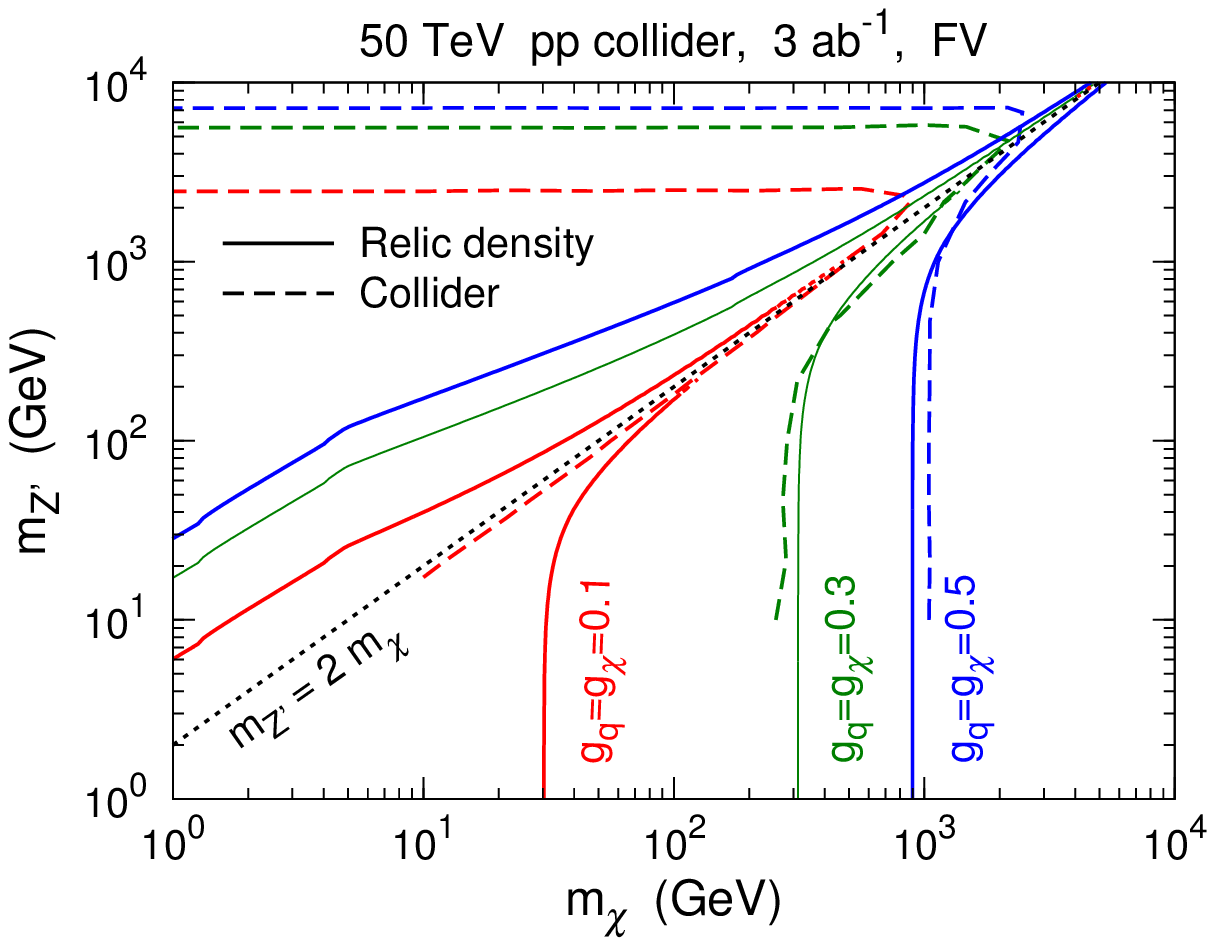}}
\subfigure[~FA model]{
\includegraphics[width=0.45\textwidth]{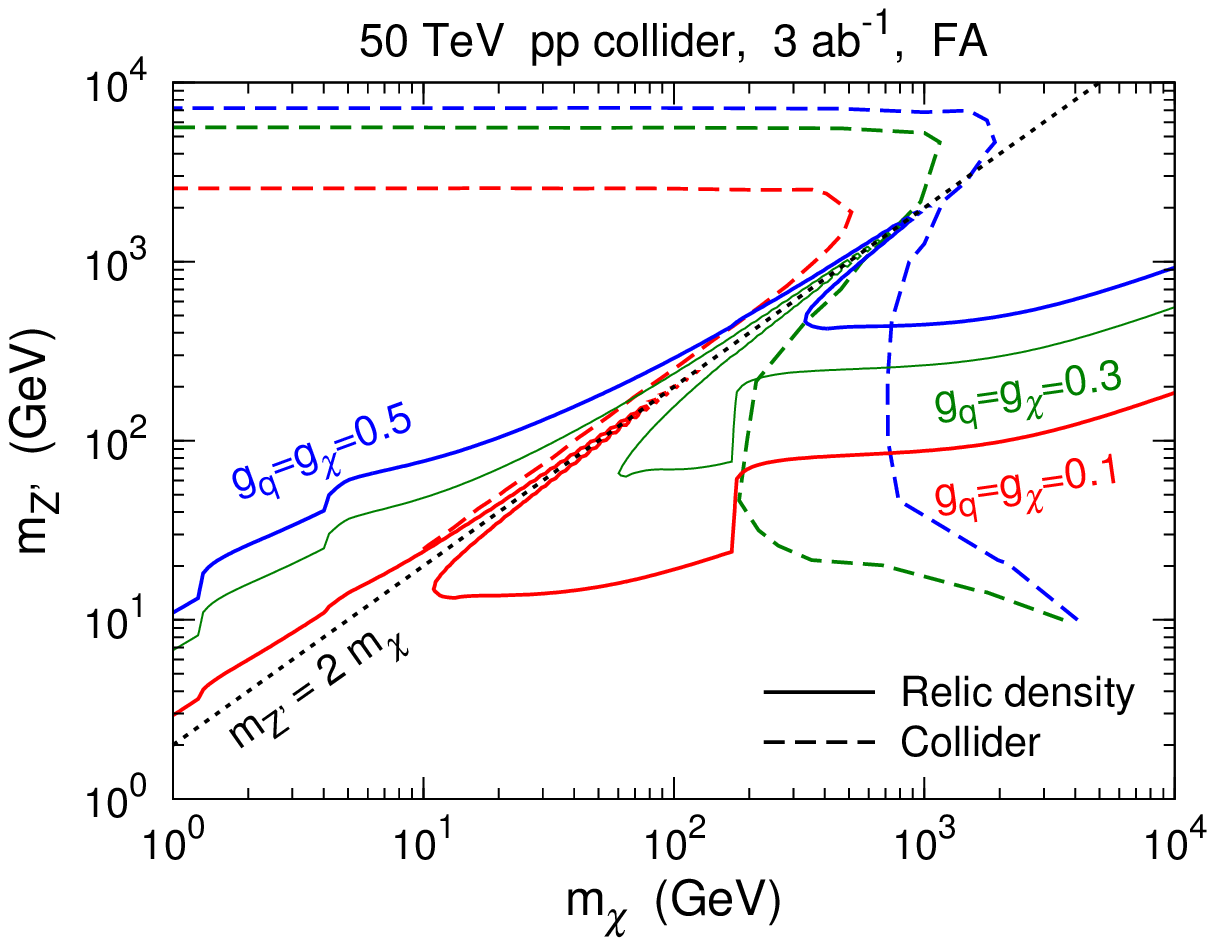}}
\subfigure[~SV model]{
\includegraphics[width=0.45\textwidth]{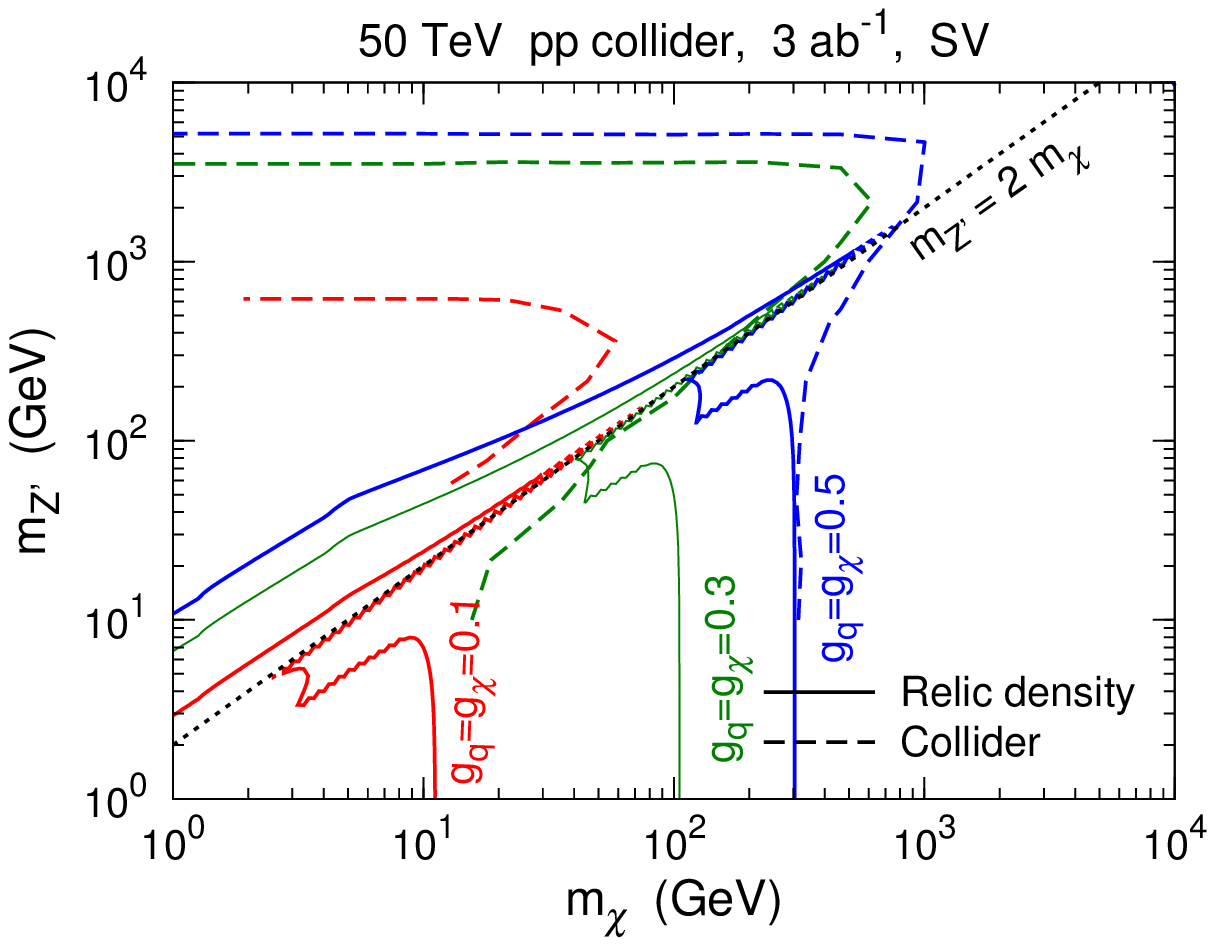}}
\caption{Estimated 90\% C.L. limits (dashed lines) on the FV~(a), FA~(b), and SV~(c) models in the $m_{Z'}$-$m_\chi$ plane for the
$\mathrm{monojet}+\missET$ channel at a $pp$ collider with $\sqrt{s} = 50~\TeV$, in comparison with the regions allowed by the observed DM thermal relic density.
Red/green/blue lines correspond to $g_q = g_\chi =0.1/0.3/0.5$.
Regions enclosed by solid curves predict a relic density smaller than the Planck+WMAP observation $\Omega_\chi h^2 = 0.1173$~\cite{Ade:2014zfo}.}
\label{fig:Omega}
\end{figure}

Finally, let us carefully inspect the connection between the collider searches and the DM relic density.
The observed DM relic density recently given by the Planck collaboration is
$\Omega_\chi h^2 = 0.1173\pm 0.0031$~\cite{Ade:2014zfo}.
Assuming DM is thermally produced, the regions with small annihilation cross section may overproduce DM in the early Universe and thus be excluded. On the other hand, the regions with DM underproduction may still survive in some situations, for instance, there exist other DM components, or DM particles are non-thermally produced from decays of new heavier particles.

We show in Fig.~\ref{fig:Omega} the regions where the estimated thermal DM relic density is smaller than the observed value.
In the FV model, the monojet search at a $pp$ collider with $\sqrt{s}=50~\TeV$ is expected to explore most of the allowed regions for $g_q=g_\chi\geq 0.3$.
In the FA model, some allowed regions with $m_\chi>m_t$ cannot be tested by the collider search.
When estimating the relic density in the SV model, we also include the annihilation contribution from the interaction term $g_\chi^2 \chi^*\chi Z'_\mu Z'^\mu$, which is naturally induced in the $U(1)$ gauge invariant extension of the SV model. This contribution can be significant for $m_\chi> m_{Z'}$.
For $g_q=g_\chi= 0.5$, the monojet search can cover the entire region allowed by the relic density observation in the SV model.

If there are some other interactions that are not included in our simplified models, the thermal DM annihilation can be enhanced and reduce the overproduction of DM particles. In this case, the monojet search may not be affected by the additional interactions, but the regions allowed by the observed relic density would be enlarged.
For instance, there may exist some $Z'$-lepton interactions which enhance DM annihilations without affecting DM productions at $pp$ colliders. Other possibilities may arise from the UV extension of the simplified models.
If the mass of $Z'$ and $\chi$ are obtained via a spontaneously symmetry breaking mechanism, an additional Higgs boson $h'$ may exist and mediate a new annihilation process $\chi \bar{\chi}/\chi \chi^* \rightarrow h' \rightarrow Z'Z'$.
In this case, the correct relic density may be more easily achieved.

\section{Conclusions and discussions}

In this work, we consider a class of DM simplified models and investigate the prospect of future $pp$ colliders.
These models contain a new vector boson connecting DM particles to quarks via vector or axial vector interactions.
In particular, we study the monojet channel at future colliders with collision energies of 33, 50, and 100~TeV.

In the simplified models, full kinematics and topologies can be well investigated. The DM pair-production cross section depends on the DM particle mass, the mediator mass and the mediator couplings.
Future colliders have capabilities to search for the mediators
with masses of $\mathcal{O}(\TeV)$.
If $m_\chi < m_{Z'}/2$, the mediator is on-shell produced and then decays into DM particles. In this case, the DM production rate is resonantly enhanced. On the other hand, if $m_\chi > m_{Z'}/2$ the sensitivity would drop quickly since the mediator is off-shell produced.
As a result, future colliders may be more sensitive to a heavier on-shell mediator than a lighter off-shell mediator.

We also compare our expectation of the collider sensitivity with constraints from the direct and indirect DM detection.
Although the collider phenomenologies of vector and axial vector interactions are similar, signal features in other DM detection
experiments can be quite different. For the vector interaction, collider searches have stronger capability to detect light DM with masses smaller than $\sim 10~\GeV$. On the other hand, the collider sensitivity to the axial vector interaction can be better than direct searches by several orders of magnitude. Moreover, collider searches would be much more sensitive than indirect searches for DM particles with masses smaller than the mediator mass. Furthermore, future collider searches could explore the bulk of the parameter space that is allowed by the observed DM relic density.

Future $pp$ colliders with very high collision energies will be sensitive to very heavy particles in new physics models. In general, other energetic objects,
such as charged leptons, photons, and $b$-jets, can be used to trigger DM signatures. With a very high luminosity, one can even utilize reconstructed gauge bosons, Higgs bosons, and top quarks to search for DM particles. These kinds of searches may benefit from small SM backgrounds at a high energy scale. The challenge is how to accurately measure objects with large $p_T$ of $\sim \mathcal{O}$(TeV) and how to efficiently reconstruct particles that are highly boosted. The DM study on these collider signatures will be a great complement to the monojet search and other DM detection experiments.

\begin{acknowledgments}
This work is supported by the National Natural Science Foundation of China
under Grants NO. 11475189, 11475191, 11135009, 11175251,
and the 973 Program of China under Grant No. 2013CB837000.
\end{acknowledgments}

\appendix

\section{DM-nucleus scattering cross sections}

Direct detection experiments measure the nuclear recoil energy induced by the DM-nucleus scattering, where the momentum transfer is far smaller than the reduced mass of the DM-nucleus system. In the limit of zero momentum transfer, the mediator in the DM-quark scattering can be safely integrated out. We can use effective Lagrangians to describe DM-quark interactions as
\begin{eqnarray}
  \mathcal{L}_\mathrm{FV}^\mathrm{eff} &  =  & - \sum_q G_{\chi q}
      \bar{\chi} \gamma^\mu \chi  \bar{q} \gamma_\mu q \label{eq:EFT_v} , \\
  \mathcal{L}_\mathrm{FA}^\mathrm{eff} &  =  & - \sum_q G_{\chi q}
      \bar{\chi} \gamma^\mu \gamma^5 \chi  \bar{q} \gamma_\mu \gamma^5 q ,  \label{eq:EFT_a}\\
  \mathcal{L}_\mathrm{SV}^\mathrm{eff} &  =  & - \sum_q G_{\chi q}
      [\chi^* \partial^\mu \chi -(\partial^\mu \chi^*) \chi]  \bar{q} \gamma_\mu q ,
    \label{eq:EFT_s}
\end{eqnarray}
where $G_{\chi q} \equiv g_q g_\chi /m_{Z'}^2$ is the effective DM-quark coupling. The vector interactions \eqref{eq:EFT_v} and \eqref{eq:EFT_s} lead to SI scatterings, while the axial vector interaction \eqref{eq:EFT_a} leads to SD scatterings.

DM-quark interactions induce DM-nucleus interactions. The DM-nucleus scattering cross sections $\sigma_{\chi A} $
are given by
\begin{eqnarray}
  \sigma_{\mathrm{SI},\,\chi A} & = & \frac{m_\chi^2 m_A^2}{\pi (m_\chi +m_A)^2 } \left[ Z G_{\mathrm{V},\,\chi p} + (A-Z) G_{\mathrm{V},\,\chi n}\right]^2
\quad\text{for the FV and SV models},
  \\
  \sigma_{\mathrm{SD},\,\chi A} & = & \frac{3 m_\chi^2 m_A^2}{\pi (m_\chi +m_A)^2 }  \left( S^A_p G_{\mathrm{A},\,\chi p} + S_n^A G_{\mathrm{A},\,\chi n}\right)^2
  \quad\text{for the FA model}.
\end{eqnarray}
Here $m_A$ is the nucleus mass.
$Z$ and $A$ are the charge number and mass number of the nucleus, respectively.
$S_p^A$ ($S_n^A$) is the expectation value
of the spin contributed by protons (neutrons) in the nucleus.
The effective DM-proton and DM-neutron couplings $G_{\chi p}$ and $G_{\chi n}$ are given by
\begin{eqnarray}
  G_{\mathrm{V},\,\chi p} &=&  2 G_{\chi u} + G_{\chi d},\quad
  G_{\mathrm{V},\,\chi n} = G_{\chi u} + 2 G_{\chi d}; \\
  G_{\mathrm{A},\,\chi p} &=& \sum_{q=u,d,s} G_{\chi q} \Delta_q^p,\quad
  G_{\mathrm{A},\,\chi n} = \sum_{q=u,d,s} G_{\chi q} \Delta_q^n,
\end{eqnarray}
where the nucleon form factors for the axial vector interaction are $\Delta_d^n =\Delta_u^p = 0.842 \pm 0.012$, $\Delta_u^n =\Delta_d^p = -0.427 \pm 0.013$,
and $\Delta_s^n =\Delta_s^p = -0.085 \pm 0.018$~\cite{Airapetian:2006vy}.

\section{DM annihilation cross sections and relic density}

DM annihilations could induce high energy cosmic rays, gamma rays, and neutrinos, which may be observed by indirect detection experiments.
The DM annihilation cross sections into quarks in the simplified models are given by
\begin{eqnarray}
  \sigma_\mathrm{FV,\,ann} (\chi\bar\chi\to q\bar{q}) & = & \sum_q \frac{\beta_q }{12 \pi \beta_\chi}
                 \frac{c_q  (g_q g_\chi)^2 s }{(s-m_{Z'}^2)^2 +m_{Z'}^2\Gamma_{Z'}^2 }
		 \left(1+ \frac{2 m_q^2}{s}\right)\left(1+ \frac{2 m_\chi^2}{s}\right), \label{eq:sv_v} \\
  \sigma_\mathrm{FA,\,ann} (\chi\bar\chi\to q\bar{q}) & = & \sum_q \frac{\beta_q }{48 \pi \beta_\chi}
                 \frac{c_q  (g_q g_\chi)^2 s }{(s-m_{Z'}^2)^2 +m_{Z'}^2\Gamma_{Z'}^2 } \nonumber\\
		&& \times \left(3+\beta_\chi^2 \beta_q^2 -12\frac{m_\chi^2}{s}-12\frac{m_q^2}{s}
		 +96 \frac{m_q^2 m_\chi^2}{s^2} -96 \frac{m_q^2 m_\chi^2}{s m_{Z'}^2}
		 +48 \frac{m_q^2 m_\chi^2}{m_{Z'}^4}\right), \nonumber\\
\\
  \sigma_\mathrm{SV,\,ann} (\chi\chi^*\to q\bar{q}) & = & \sum_q \frac{\beta_q \beta_\chi}{12 \pi}
               \frac{c_q  (g_q g_\chi)^2 s }{
               (s-m_{Z'}^2)^2 +m_{Z'}^2\Gamma_{Z'}^2 }\left(1+
	         \frac{2 m_q^2}{s}\right), \label{eq:sv_s}
\end{eqnarray}
where $s$ is the squared center-of-mass energy of a DM particle pair, $\beta_\chi\equiv\sqrt{1-4 m_\chi^2/s}$, and $\beta_q\equiv\sqrt{1-4m_q^2/s}$.
DM particles can also annihilate into $Z'$ pairs via $t$-channel and $u$-channel if $m_\chi>m_{Z'}$. For the SV model, we also consider the contribution from the interaction term $g_\chi^2 \chi^*\chi Z'_\mu Z'^\mu$ introduced in a U(1) gauge extension of the simplified model.

Taking into account the velocity distribution of DM particles,
the quantity directly connecting to indirect searches is the thermally averaged annihilation cross section $ \langle \sigma_\mathrm{ann} v_M\rangle$, where $v_M\equiv \sqrt{(p_1\cdot p_2)^2-m_1^2 m_2^2}/(E_1E_2)$ is the M${\o}$ller velocity.
As pointed in Ref.~\cite{Gondolo:1990dk}, instead of directly calculating $ \langle
\sigma_\mathrm{ann} v_M\rangle$, one can conveniently calculate $\langle
\sigma_\mathrm{ann} v_\mathrm{rel}\rangle$ in the laboratory frame and give the same result.
Here the laboratory frame means one of the two initial particles is at rest, and $v_\mathrm{rel}$ is the relative velocity between them.

In the laboratory frame with the low DM velocity, $s$ can be expanded as
$s = 4 m_\chi^2 + m_\chi^2 v^2 +\frac{3}{4} m_\chi^2 v^4 +\mathcal{O}(v^6)$,
where $v\equiv v_\mathrm{rel}=\beta_\chi (1-2m_\chi^2/s)^{-1}$.
Substituting this expression into $\sigma_\mathrm{ann}$,
we can expand $\sigma_\mathrm{ann} v$ as $a+ b v^2 + \mathcal{O}(v^4)$.
For a Maxwell-Boltzmann velocity distribution parametrized by temperature $T$,
we have $ \langle\sigma_\mathrm{ann} v \rangle = a + 6 b x^{-1} +\mathcal{O}(x^{-2})$, where $x\equiv m_\chi/T$.
We list below the coefficients $a$ and $b$ in the simplified models.
In the FV model,
\begin{eqnarray}
    a_\mathrm{FV}&=& \frac{g_\chi^4 (1-m_{Z'}^2/m_\chi^2)^{3/2} }{4 \pi(m_{Z'}^2-2 m_\chi^2)^2}m_\chi^2 \Theta(m_\chi-m_{Z'}) \nonumber \\
      &+&\sum_q \frac{c_q g_q^2 g_\chi^2
      \sqrt{1-m_q^2/m_\chi^2} } {2\pi [(m_{Z'}^2-4 m_\chi^2)^2
      +m_{Z'}^2\Gamma_{Z'}^2]} (2m_\chi^2+m_q^2),\\
      b_\mathrm{FV}&=& \frac{g_\chi^4 \sqrt{1-m_{Z'}^2/m_\chi^2} }{96 \pi (m_{Z'}^2-2 m_\chi^2 )^4}
    (23 m_{Z'}^6 -66 m_\chi^2 m_{Z'}^4 +76 m_\chi^4 m_{Z'}^2) \Theta(m_\chi-m_{Z'})
\nonumber\\
   &+& \sum_q \frac{c_q g_q^2 g_\chi^2 } {48 \pi m_\chi^2
	\sqrt{1-m_q^2/m_\chi^2}[(m_{Z'}^2-4 m_\chi^2)^2
	+m_{Z'}^2\Gamma_{Z'}^2]^2} \{m_{Z'}^2\Gamma_{Z'}^2(2m_q^2 m_\chi^2
	+11m_q^4 -4 m_\chi^2) \nonumber\\
	&& +\left( m_{Z'}^2-4 m_\chi^2
	\right)[-4m_\chi^4(14m_q^2+m_{Z'}^2)+2m_q^2m_\chi^2
	(m_{Z'}^2-46m_q^2)+11m_q^4m_{Z'}^2+112m_\chi^6]\}, \nonumber \\
\end{eqnarray}
where the step function means the annihilation channel $\chi \bar{\chi} \rightarrow Z'Z'$ opens only if $m_\chi>m_{Z'}$. In the FA model,
\begin{eqnarray}
    a_\mathrm{FA}&=& \frac{g_\chi^4 (1-m_{Z'}^2/m_\chi^2)^{3/2} }{4 \pi(m_{Z'}^2-2 m_\chi^2)^2} m_\chi^2 \Theta(m_\chi-m_{Z'}) \nonumber \\
    &+& \sum_q \frac{c_q g_q^2 g_\chi^2
      \sqrt{1-m_q^2/m_\chi^2}} {2\pi [(m_{Z'}^2-4 m_\chi^2)^2
      +m_{Z'}^2\Gamma_{Z'}^2]} m_q^2 \left(1-\frac{4m_\chi^2}{m_{Z'}^2}\right)^2,\\
    b_\mathrm{FA}&=& \frac{g_\chi^4 \sqrt{1-m_{Z'}^2/m_\chi^2} }{96 \pi m_{Z'}^4 (m_{Z'}^2-2 m_\chi^2 )^4}
\Theta(m_\chi-m_{Z'}) \nonumber\\
&&\times (23  m_{Z'}^{10} -118m_\chi^2 m_{Z'}^8+172 m_\chi^4 m_{Z'}^6 +32m_\chi^6 m_{Z'}^4 -192 m_\chi^8 m_{Z'}^2+128 m_\chi^{10})
\nonumber\\
    && + \sum_q \frac{c_q
      g_q^2 g_\chi^2 } {48 \pi m_\chi^2 m_{Z'}^2
      \sqrt{1-m_q^2/m_\chi^2}[(m_{Z'}^2-4 m_\chi^2)^2
      +m_{Z'}^2\Gamma_{Z'}^2]^2} \nonumber\\
    &&\quad\times \{m_{Z'}^2\Gamma_{Z'}^2[-4 m_q^2
      m_\chi^2 m_{Z'}^2 (18m_q^2+7m_{Z'}^2) + 8m_\chi^4(6m_q^2 m_{Z'}^2 +6
    m_q^4 +m_{Z'}^4 ) +23 m_q^4 m_{Z'}^4 ] \nonumber\\
    &&~\qquad +\left( m_{Z'}^2-4 m_\chi^2
    \right)^2 [m_q^4(240 m_\chi^4 -120 m_\chi^2 m_{Z'}^2 +23 m_{Z'}^4) \nonumber\\
    &&~\qquad\qquad\qquad\qquad\qquad -4m_q^2(48 m_\chi^6 -24 m_\chi^4 m_{Z'}^2 +7 m_\chi^2 m_{Z'}^4) +8m_\chi^4
  m_{Z'}^4 ] \}.
\end{eqnarray}
$a_\mathrm{FA}$ for DM annihilations into quarks are proportional to $m_q^2$, implying that the annihilations are helicity suppressed in $s$-wave.
In the SV model,
\begin{eqnarray}
    a_\mathrm{SV}&=& \frac{g_\chi^4 \sqrt{1-m_{Z'}^2/m_\chi^2} }{16 \pi(m_{Z'}^2-2 m_\chi^2)^2}
    (8 m_\chi^4-8 m_\chi^2 m_{Z'}^2 +3 m_{Z'}^4) \Theta(m_\chi-m_{Z'}),  \\
    b_\mathrm{SV}&=& \frac{g_\chi^4}{384 \pi (m_{Z'}^2-2m_\chi^2)^4 \sqrt{1-m_{Z'}^2/m_\chi^2}} \Theta(m_\chi-m_{Z'})
\nonumber\\
      && \times (-640 m_\chi^{10} +1888 m_\chi^8 m_{Z'}^2 -2224 m_\chi^6 m_{Z'}^4 +1332 m_\chi^4 m_{Z'}^6
      -392 m_\chi^2 m_{Z'}^8 +45 m_{Z'}^{10})
\nonumber\\
     &&+\sum_q \frac{c_q g_q^2 g_\chi^2
      \sqrt{1-m_q^2/m_\chi^2} } {12 \pi [(m_{Z'}^2-4 m_\chi^2)^2
      +m_{Z'}^2\Gamma_{Z'}^2]} (2m_\chi^2+m_q^2).
\end{eqnarray}
DM annihilations into quarks are of $p$-wave in the leading order, because a pair of scalar particles cannot form a vector state without orbital angular momentum.

The evolution of thermal DM density can be determined by the Boltzmann equation
 \begin{equation}
   \frac{d n_\chi}{dt} +3 H n_\chi
   = -\left<\sigma_\mathrm{ann} v\right> (n_\chi n_{\bar{\chi}}-  n_\chi^\mathrm{eq} n_{\bar{\chi}}^\mathrm{eq})
   = -\left<\sigma_\mathrm{ann} v\right> [(n_\chi)^2
   -  (n_\chi^\mathrm{eq})^2], \label{eq:relic}
 \end{equation}
 where $H \equiv  \sqrt{8 \pi \rho/(3 M_\mathrm{pl}^2)}$
 is the Hubble rate, $\rho$ is the energy density in the Universe, and $ M_\mathrm{pl}$ is the Planck mass. $n_\chi$ $(n_{\bar{\chi}})$ is the number density of the DM particle (antiparticle).
 The superscript ``eq'' represents the thermal equilibrium. Here we do not consider DM particle-antiparticle asymmetry and assume $n_\chi = n_{\bar{\chi}}$.

In principle, one can numerically solve the Boltzmann equation and get the DM
relic density. Here we use an approximate method to obtain~\cite{Kolb:1990vq,Jungman:1995df}
\begin{equation}
  \Omega_\chi h^2 =2 \times 1.04\times 10^9
  ~\GeV^{-1} \left(\frac{T_0}{2.725 K}\right)^3 \frac{x_f}{M_\mathrm{pl}\sqrt{g_\star (x_f)}(a+ 3
b/x_f)},
  \label{eq:O}
\end{equation}
where $x_f \equiv m_\chi /T_f\sim \mathcal{O}(10)$,
$T_f$ is the DM freeze-out temperature,
$T_0= 2.725\pm 0.002~\mathrm{K}$~\cite{Mather:1998gm} is the present CMB temperature,
and $g_\star (x_f)$ is the effectively relativistic degrees of freedom at the freeze-out epoch~\cite{Coleman:2003hs}.

\section{Unitarity bound\label{app:uni}}

The unitarity condition for a 2-body inelastic process can be expressed as
\begin{equation}
  |a_j^\mathrm{inel}(\hat{s})| \le \frac{1}{2 \sqrt{\beta(\hat{s},m_\mathrm{in}) \beta(\hat{s},m_\mathrm{out})}},
\end{equation}
where $\beta(\hat{s},m)\equiv\sqrt{1-4m^2/\hat{s}}$, $m_\mathrm{in}$ ($m_\mathrm{out}$) is the mass of either of the two incoming (outgoing) particles, and $\hat{s}$ is the center-of-mass energy of the system.
$ a_j^\mathrm{inel}(\hat{s})$ is the $j$-th partial wave coefficient of the invariant amplitude $\mathcal{M}$ for the process:
\begin{equation}
  a_j^\mathrm{inel}(\hat{s}) = \frac{1}{32 \pi} \int_0^\pi d\theta \sin \theta
  P_j(\cos \theta) \mathcal{M}^\mathrm{inel}(\hat{s},\cos \theta),
\end{equation}
where $P_j(x)$ is the $j$-th Legendre polynomial.

In the FA model, the 0-th partial wave coefficient for the process $q_-\bar{q}_-\to\chi_-\bar{\chi}_-$ (the minus sign means a helicity of $-1$) is given by
\begin{equation}
a_0^\mathrm{inel}(\hat{s}) = \frac{g_q g_\chi}{4 \pi} \frac{m_q m_\chi}{\hat{s}-m_{Z'}^2+i m_{Z'} \Gamma_{Z'}} \left(1-\frac{\hat{s}}{m_{Z'}^2}\right).
\label{eq:a0}
\end{equation}
Apparently, this process violates the unitarity condition when $m_{Z'}\ll \sqrt{\hat{s}}$.

In order to obtain the unitarity bound on DM particle pair productions at $pp$ colliers, one challenge is how to determine $\hat{s}$,
which varies event by event and is affected by parton distribution functions and associated jets. There are several estimates of $\hat{s}$ used in the literature,
such as the center-of-mass energy of the collider~\cite{Shoemaker:2011vi},
$\sqrt{4 m_\chi^2 + p_T^2}$~\cite{Buchmueller:2013dya,Buchmueller:2014yoa},
and an estimate of the invariant mass of the DM particle pair in some fraction of events~\cite{Fox:2012ee}.
In this work, we adopt the strategy in Ref.~\cite{Busoni:2013lha} and take $\hat{s}$ as the averaged momentum transfer weighting with parton distribution functions:
\begin{equation}
\hat{s} = \left<Q^2 \right>= \frac{\sum_q \int dx_1 dx_2 [f_q(x_1)f_{\bar{q}}(x_2)+
  f_q(x_2)f_{\bar{q}}(x_1)]\Theta (Q-2 m_\chi) Q^2}{
    \sum_q \int dx_1 dx_2 [f_q(x_1)f_{\bar{q}}(x_2)+
    f_q(x_2)f_{\bar{q}}(x_1)]\Theta (Q-2 m_\chi)},
\end{equation}
where $Q^2 = (p_q+p_{\bar{q}}-p_j)^2 =x_1 x_2 s -\sqrt{s} \pT (x_1 e^{-\eta}+x_2 e^\eta)$ is the squared momentum transfer to the DM particle pair in the process $q\bar{q}\to\chi\bar\chi j$.
$\pT$ and $\eta$ are the transverse momentum and
pseudo-rapidity of the jet $j$, respectively.
The unitarity bound demonstrated in Fig.~\ref{fig:50TeV_collider} is derived from the process $b_-\bar{b}_-\to\chi_-\bar{\chi}_-$. We have set $\eta=0$
and $\pT$ as the cut threshold in order to give a conservative bound.


\begin{thebibliography}{99}

\bibitem{Beltran:2010ww}
  M.~Beltran, D.~Hooper, E.~W.~Kolb, Z.~A.~C.~Krusberg and T.~M.~P.~Tait,
  JHEP {\bf 1009}, 037 (2010)
  [arXiv:1002.4137 [hep-ph]].

\bibitem{Fan:2010gt}
  J.~Fan, M.~Reece and L.~T.~Wang,
  JCAP {\bf 1011}, 042 (2010)
  [arXiv:1008.1591 [hep-ph]].



\bibitem{Cao:2009uw}
  Q.~H.~Cao, C.~R.~Chen, C.~S.~Li and H.~Zhang,
  JHEP {\bf 1108}, 018 (2011)
  [arXiv:0912.4511 [hep-ph]].



\bibitem{Goodman:2010yf}
  J.~Goodman, M.~Ibe, A.~Rajaraman, W.~Shepherd, T.~M.~P.~Tait and H.~B.~Yu,
  Phys.\ Lett.\ B {\bf 695}, 185 (2011)
  [arXiv:1005.1286 [hep-ph]].



\bibitem{Goodman:2010ku}
  J.~Goodman, M.~Ibe, A.~Rajaraman, W.~Shepherd, T.~M.~P.~Tait and H.~B.~Yu,
  Phys.\ Rev.\ D {\bf 82}, 116010 (2010)
  [arXiv:1008.1783 [hep-ph]].



\bibitem{Cheung:2010ua}
  K.~Cheung, P.~Y.~Tseng and T.~C.~Yuan,
  JCAP {\bf 1101}, 004 (2011)
  [arXiv:1011.2310 [hep-ph]].



\bibitem{Zheng:2010js}
  J.~M.~Zheng, Z.~H.~Yu, J.~W.~Shao, X.~J.~Bi, Z.~Li and H.~H.~Zhang,
  Nucl.\ Phys.\ B {\bf 854}, 350 (2012)
  [arXiv:1012.2022 [hep-ph]].



\bibitem{Cheung:2011nt}
  K.~Cheung, P.~Y.~Tseng and T.~C.~Yuan,
  JCAP {\bf 1106}, 023 (2011)
  [arXiv:1104.5329 [hep-ph]].



\bibitem{Yu:2011by}
  Z.~H.~Yu, J.~M.~Zheng, X.~J.~Bi, Z.~Li, D.~X.~Yao and H.~H.~Zhang,
  Nucl.\ Phys.\ B {\bf 860}, 115 (2012)
  [arXiv:1112.6052 [hep-ph]].



\bibitem{Fitzpatrick:2012ix}
  A.~L.~Fitzpatrick, W.~Haxton, E.~Katz, N.~Lubbers and Y.~Xu,
  JCAP {\bf 1302}, 004 (2013)
  [arXiv:1203.3542 [hep-ph]].


\bibitem{Aad:2015zva}
  G.~Aad {\it et al.}  [ATLAS Collaboration],
  arXiv:1502.01518 [hep-ex].

\bibitem{Aad:2014tda}
  G.~Aad {\it et al.}  [ATLAS Collaboration],
  Phys.\ Rev.\ D {\bf 91}, no. 1, 012008 (2015)
  [arXiv:1411.1559 [hep-ex]].

\bibitem{Khachatryan:2014rra}
  V.~Khachatryan {\it et al.}  [CMS Collaboration],
  arXiv:1408.3583 [hep-ex].

\bibitem{Khachatryan:2014rwa}
  V.~Khachatryan {\it et al.}  [CMS Collaboration],
  arXiv:1410.8812 [hep-ex].

\bibitem{Bai:2010hh}
  Y.~Bai, P.~J.~Fox and R.~Harnik,
  JHEP {\bf 1012}, 048 (2010)
  [arXiv:1005.3797 [hep-ph]].



\bibitem{Fox:2011pm}
  P.~J.~Fox, R.~Harnik, J.~Kopp and Y.~Tsai,
  Phys.\ Rev.\ D {\bf 85}, 056011 (2012)
  [arXiv:1109.4398 [hep-ph]].



\bibitem{Shoemaker:2011vi}
  I.~M.~Shoemaker and L.~Vecchi,
  Phys.\ Rev.\ D {\bf 86}, 015023 (2012)
  [arXiv:1112.5457 [hep-ph]].



\bibitem{Buchmueller:2013dya}
  O.~Buchmueller, M.~J.~Dolan and C.~McCabe,
  JHEP {\bf 1401}, 025 (2014)
  [arXiv:1308.6799 [hep-ph], arXiv:1308.6799].



\bibitem{Busoni:2013lha}
  G.~Busoni, A.~De Simone, E.~Morgante and A.~Riotto,
  Phys.\ Lett.\ B {\bf 728}, 412 (2014)
  [arXiv:1307.2253 [hep-ph]].



\bibitem{Busoni:2014sya}
  G.~Busoni, A.~De Simone, J.~Gramling, E.~Morgante and A.~Riotto,
  JCAP {\bf 1406}, 060 (2014)
  [arXiv:1402.1275 [hep-ph]].

\bibitem{Endo:2014mja}
  M.~Endo and Y.~Yamamoto,
  JHEP {\bf 1406}, 126 (2014)
  [arXiv:1403.6610 [hep-ph]].

\bibitem{Busoni:2014haa}
  G.~Busoni, A.~De Simone, T.~Jacques, E.~Morgante and A.~Riotto,
  JCAP {\bf 1409}, 022 (2014)
  [arXiv:1405.3101 [hep-ph]].

\bibitem{Buchmueller:2014yoa}
  O.~Buchmueller, M.~J.~Dolan, S.~A.~Malik and C.~McCabe,
  JHEP {\bf 1501}, 037 (2015)
  [arXiv:1407.8257 [hep-ph]].


\bibitem{Cohen:2013xda}
  T.~Cohen, T.~Golling, M.~Hance, A.~Henrichs, K.~Howe, J.~Loyal, S.~Padhi and J.~G.~Wacker,
  JHEP {\bf 1404}, 117 (2014)
  [arXiv:1311.6480 [hep-ph]].

\bibitem{Kraml:2013mwa}
  S.~Kraml, S.~Kulkarni, U.~Laa, A.~Lessa, W.~Magerl, D.~Proschofsky-Spindler and W.~Waltenberger,
  Eur.\ Phys.\ J.\ C {\bf 74}, 2868 (2014)
  [arXiv:1312.4175 [hep-ph]].

\bibitem{Papucci:2014rja}
  M.~Papucci, K.~Sakurai, A.~Weiler and L.~Zeune,
  Eur.\ Phys.\ J.\ C {\bf 74}, no. 11, 3163 (2014)
  [arXiv:1402.0492 [hep-ph]].

\bibitem{Lebedev:2014bba}
  O.~Lebedev and Y.~Mambrini,
  Phys.\ Lett.\ B {\bf 734}, 350 (2014)
  [arXiv:1403.4837 [hep-ph]].


\bibitem{Abdallah:2014hon}
  J.~Abdallah, A.~Ashkenazi, A.~Boveia, G.~Busoni, A.~De Simone, C.~Doglioni, A.~Efrati and E.~Etzion {\it et al.},
  arXiv:1409.2893 [hep-ph].



\bibitem{Berlin:2014tja}
  A.~Berlin, D.~Hooper and S.~D.~McDermott,
  Phys.\ Rev.\ D {\bf 89}, no. 11, 115022 (2014)
  [arXiv:1404.0022 [hep-ph]].

\bibitem{Yu:2014mfa}
  Z.~H.~Yu, X.~J.~Bi, Q.~S.~Yan and P.~F.~Yin,
  Phys.\ Rev.\ D {\bf 91}, no. 3, 035008 (2015)
  [arXiv:1410.3347 [hep-ph]].


\bibitem{Busoni:2014gta}
  G.~Busoni, A.~De Simone, T.~Jacques, E.~Morgante and A.~Riotto,
  arXiv:1410.7409 [hep-ph].


\bibitem{Hooper:2014fda}
  D.~Hooper,
  Phys.\ Rev.\ D {\bf 91}, no. 3, 035025 (2015)
  [arXiv:1411.4079 [hep-ph]].

\bibitem{Alves:2015pea}
  A.~Alves, A.~Berlin, S.~Profumo and F.~S.~Queiroz,
  arXiv:1501.03490 [hep-ph].

\bibitem{Chen:2015tia}
  N.~Chen, J.~Wang and X.~P.~Wang,
  arXiv:1501.04486 [hep-ph].

\bibitem{Anderson:2013ida}
  J.~Anderson,
  arXiv:1309.0845 [hep-ex].



\bibitem{Fowlie:2014awa}
  A.~Fowlie and M.~Raidal,
  Eur.\ Phys.\ J.\ C {\bf 74}, 2948 (2014)
  [arXiv:1402.5419 [hep-ph]].



\bibitem{Wen:2014mha}
  Y.~Wen, H.~Qu, D.~Yang, Q.~s.~Yan, Q.~Li and Y.~Mao,
  arXiv:1407.4922 [hep-ph].



\bibitem{Anchordoqui:2014wha}
  L.~A.~Anchordoqui, I.~Antoniadis, D.~C.~Dai, W.~Z.~Feng, H.~Goldberg, X.~Huang, D.~Lust and D.~Stojkovic {\it et al.},
  Phys.\ Rev.\ D {\bf 90}, no. 6, 066013 (2014)
  [arXiv:1407.8120 [hep-ph]].



\bibitem{Alva:2014gxa}
  D.~Alva, T.~Han and R.~Ruiz,
  JHEP {\bf 1502}, 072 (2015)
  [arXiv:1411.7305 [hep-ph]].



\bibitem{An:2012va}
  H.~An, X.~Ji and L.~T.~Wang,
  JHEP {\bf 1207}, 182 (2012)
  [arXiv:1202.2894 [hep-ph]].

\bibitem{Yu:2013wta}
  F.~Yu,
  arXiv:1308.1077 [hep-ph].

\bibitem{Buckley:2014fba}
  M.~R.~Buckley, D.~Feld and D.~Goncalves,
  Phys.\ Rev.\ D {\bf 91}, no. 1, 015017 (2015)
  [arXiv:1410.6497 [hep-ph]].



\bibitem{Alwall:2014hca}
  J.~Alwall, R.~Frederix, S.~Frixione, V.~Hirschi, F.~Maltoni, O.~Mattelaer, H.-S.~Shao and T.~Stelzer {\it et al.},
  JHEP {\bf 1407}, 079 (2014)
  [arXiv:1405.0301 [hep-ph]].

\bibitem{Alloul:2013bka}
  A.~Alloul, N.~D.~Christensen, C.~Degrande, C.~Duhr and B.~Fuks,
  Comput.\ Phys.\ Commun.\  {\bf 185}, 2250 (2014)
  [arXiv:1310.1921 [hep-ph]].

\bibitem{Sjostrand:2006za}
  T.~Sjostrand, S.~Mrenna and P.~Z.~Skands,
  JHEP {\bf 0605}, 026 (2006)
  [hep-ph/0603175].

\bibitem{deFavereau:2013fsa}
  J.~de Favereau {\it et al.}  [DELPHES 3 Collaboration],
  JHEP {\bf 1402}, 057 (2014)
  [arXiv:1307.6346 [hep-ex]].

\bibitem{Cacciari:2008gp}
  M.~Cacciari, G.~P.~Salam and G.~Soyez,
  JHEP {\bf 0804}, 063 (2008)
  [arXiv:0802.1189 [hep-ph]].




\bibitem{Aprile:2012nq}
  E.~Aprile {\it et al.}  [XENON100 Collaboration],
  Phys.\ Rev.\ Lett.\  {\bf 109}, 181301 (2012)
  [arXiv:1207.5988 [astro-ph.CO]].



\bibitem{Akerib:2013tjd}
  D.~S.~Akerib {\it et al.}  [LUX Collaboration],
  Phys.\ Rev.\ Lett.\  {\bf 112}, 091303 (2014)
  [arXiv:1310.8214 [astro-ph.CO]].



\bibitem{Agnese:2014aze}
  R.~Agnese {\it et al.}  [SuperCDMS Collaboration],
  Phys.\ Rev.\ Lett.\  {\bf 112}, no. 24, 241302 (2014)
  [arXiv:1402.7137 [hep-ex]].



\bibitem{Aprile:2012zx}
  E.~Aprile [XENON1T Collaboration],
  Springer Proc.\ Phys.\  {\bf 148}, 93 (2013)
  [arXiv:1206.6288 [astro-ph.IM]].



\bibitem{Felizardo:2011uw}
  M.~Felizardo, T.~A.~Girard, T.~Morlat, A.~C.~Fernandes, A.~R.~Ramos, J.~G.~Marques, A.~Kling and J.~Puibasset {\it et al.},
  Phys.\ Rev.\ Lett.\  {\bf 108}, 201302 (2012)
  [arXiv:1106.3014 [astro-ph.CO]].



\bibitem{Archambault:2012pm}
  S.~Archambault {\it et al.}  [PICASSO Collaboration],
  Phys.\ Lett.\ B {\bf 711}, 153 (2012)
  [arXiv:1202.1240 [hep-ex]].



\bibitem{Behnke:2012ys}
  E.~Behnke {\it et al.}  [COUPP Collaboration],
  Phys.\ Rev.\ D {\bf 86}, no. 5, 052001 (2012)
  [Erratum-ibid.\ D {\bf 90}, no. 7, 079902 (2014)]
  [arXiv:1204.3094 [astro-ph.CO]].



\bibitem{Tanaka:2011uf}
  T.~Tanaka {\it et al.}  [Super-Kamiokande Collaboration],
  Astrophys.\ J.\  {\bf 742}, 78 (2011)
  [arXiv:1108.3384 [astro-ph.HE]].



\bibitem{IceCube:2011aj}
  R.~Abbasi {\it et al.}  [IceCube Collaboration],
  Phys.\ Rev.\ D {\bf 85}, 042002 (2012)
  [arXiv:1112.1840 [astro-ph.HE]].



\bibitem{Ackermann:2013yva}
  M.~Ackermann {\it et al.}  [Fermi-LAT Collaboration],
  Phys.\ Rev.\ D {\bf 89}, 042001 (2014)
  [arXiv:1310.0828 [astro-ph.HE]].

\bibitem{Doro:2012xx}
  M.~Doro {\it et al.}  [CTA Collaboration],
  Astropart.\ Phys.\  {\bf 43}, 189 (2013)
  [arXiv:1208.5356 [astro-ph.IM]].

\bibitem{Ade:2014zfo}
  P.~A.~R.~Ade {\it et al.}  [Planck Collaboration],
  arXiv:1406.7482 [astro-ph.CO].

\bibitem{Airapetian:2006vy}
  A.~Airapetian {\it et al.}  [HERMES Collaboration],
  Phys.\ Rev.\ D {\bf 75}, 012007 (2007)
  [hep-ex/0609039].



\bibitem{Gondolo:1990dk}
  P.~Gondolo and G.~Gelmini,
  Nucl.\ Phys.\ B {\bf 360}, 145 (1991).

\bibitem{Kolb:1990vq}
  E.~W.~Kolb and M.~S.~Turner,
  Front.\ Phys.\  {\bf 69}, 1 (1990).



\bibitem{Jungman:1995df}
  G.~Jungman, M.~Kamionkowski and K.~Griest,
  Phys.\ Rept.\  {\bf 267}, 195 (1996)
  [hep-ph/9506380].



\bibitem{Mather:1998gm}
  J.~C.~Mather, D.~J.~Fixsen, R.~A.~Shafer, C.~Mosier and D.~T.~Wilkinson,
  Astrophys.\ J.\  {\bf 512}, 511 (1999)
  [astro-ph/9810373].



\bibitem{Coleman:2003hs}
  T.~S.~Coleman and M.~Roos,
  Phys.\ Rev.\ D {\bf 68}, 027702 (2003)
  [astro-ph/0304281].


\bibitem{Fox:2012ee}
  P.~J.~Fox, R.~Harnik, R.~Primulando and C.~T.~Yu,
  Phys.\ Rev.\ D {\bf 86}, 015010 (2012)
  [arXiv:1203.1662 [hep-ph]].


\end{thebibliography}
\end{document}